\RequirePackage{fix-cm}

\documentclass[twocolumn,epjc3]{svjour3}  

\smartqed  

\RequirePackage{graphicx}
\usepackage{siunitx}
\DeclareSIUnit{\squaredegree}{\degree\squared}
\DeclareSIUnit{\degrees}{degrees}
\usepackage{float}
\usepackage{multirow}
\usepackage{multicol}
\usepackage{booktabs}
\usepackage{array}
\newcolumntype{L}[1]{>{\raggedright\arraybackslash}p{#1}}
\newcolumntype{P}[1]{>{\raggedright\arraybackslash}p{#1}}
\RequirePackage[colorlinks,citecolor=blue,urlcolor=blue,linkcolor=blue]{hyperref}
\journalname{Eur. Phys. J. C}

\begin{document}

\title{Event Reconstruction for Radio-Based In-Ice Neutrino Detectors with Neural Posterior Estimation
}

\author{Nils Heyer\thanksref{e1,addr1}
        \and
        Christian Glaser\thanksref{addr1, addr3} 
        \and
        Thorsten Glüsenkamp\thanksref{addr1,addr2}
        \and
        Martin Ravn\thanksref{addr1}
}

\thankstext{e1}{e-mail: nils.heyer@physics.uu.se}

\institute{Dept. of Physics and Astronomy, Uppsala University, Box 516, SE-75120 Uppsala, Sweden \label{addr1}
           \and
           Dept. of Physics, TU Dortmund University, Dortmund, Germany\label{addr3}
           \and
           Oskar Klein Centre and Dept. of Physics, Stockholm University, SE-10691 Stockholm, Sweden \label{addr2}
}

\date{Received: date / Accepted: date}

\maketitle

\begin{abstract}

The detection of ultra-high-energy (UHE) neutrinos in the \SI{}{\exa\eV} range is the goal of current and future in-ice radio arrays at the South Pole and in Greenland. Here, we present a deep neural network that can reconstruct the main neutrino properties of interest from the raw waveforms recorded by the radio antennas: the neutrino direction, the energy of the particle shower induced by the neutrino interaction, and the event topology, thereby estimating the neutrino flavor. For the first time, we predict the full posterior PDF for the energy and direction reconstruction via neural posterior estimation utilizing conditional normalizing flows, enabling event-by-event uncertainty prediction. We improve over previous reconstruction algorithms and obtain a median resolution of 0.30 log(E) and 18 square degrees for a 'shallow' detector component and 0.08 log(E) and 28 square degrees for a 'deep' detector component for neutral current (NC) events at a shower energy of \SI{1}{\exa\eV}. This deep learning approach also allows us to reconstruct the more stochastic $\nu_e$ - charged current (CC) events. We quantify the impact of different antenna types and systematic uncertainties on the reconstruction and derive a goodness-of-fit score to test the compatibility of measured neutrino signals with the Monte Carlo simulations used to train the neural network.

\end{abstract}

\section{Introduction}

Cosmic neutrino detection is one of the cornerstones of multimessenger astronomy. In the last decade, the IceCube neutrino observatory at the South Pole has made measurements of the diffuse flux \cite{doi:10.1126/science.1242856} and found evidence for several sources of cosmic neutrinos \cite{doi:10.1126/science.aat2890,doi:10.1126/science.abg3395,doi:10.1126/science.adc9818}. Recently, the KM3Net telescope reported on measuring the most energetic neutrino so far in the hundred PeV range \cite{km3net}. For these advances, both experiments relied heavily on their energy and direction sensitivity. However, due to the short absorption length of visible light in ice or water, detectors relying on optical Cherenkov radiation, such as IceCube or KM3NeT, require a dense instrumentation of the detector medium and are, as such, currently limited to \SI{}{\tera\eV} and \SI{}{\peta\eV} neutrinos.

To further extend the reach of cosmic neutrino detectors into the \SI{}{\exa\eV} range in a cost-effective way and to cope with the decreasing neutrino flux with energy, sparsely instrumented in-ice radio detectors are one of the most promising approaches \cite{Barwick:2022vqt}. Here, we address one of the primary analysis tasks: reliable reconstruction capabilities, which are needed to continue the advance of in-ice radio neutrino astronomy. With a successful reconstruction, meaning correct and small uncertainty contours, the detection of UHE neutrinos can push the energy frontier of diffuse neutrino flux measurements \cite{PhysRevD.107.043019}, provide insight into the most energetic environments in the universe \cite{Fiorillo_2023}, and allow measurements of fundamental neutrino properties such as cross section \cite{cross_section} and flavor composition \cite{PhysRevD.110.023044}.

The radio technique aims at capturing short radio flashes emitted from in-ice, UHE neutrino interactions. These radio flashes are created due to a time-dependent charge imbalance in neutrino-induced showers, and they become detectable as the radiation interferes coherently when emitted on the Cherenkov cone. This mechanism is called the Askaryan effect \cite{Askaryan}, which was experimentally confirmed for particle cascades in various materials \cite{ask_slac,ask_salt,ask_anita}. The attenuation length for the ice sheets at the South Pole and in Greenland is $\mathcal O$(\SI{1}{\kilo\meter}) \cite{Barwick_Besson_Gorham_Saltzberg_2005,Greenland_att}, which allows radio antennas installed near the surface to detect signals from neutrino interactions deep in the ice.

The IceCube-Gen2 Neutrino Observatory \cite{Aartsen_2021,TDR} is the planned successor to the IceCube Neutrino Observatory. As part of the extension, more than 350 radio stations are planned to be deployed covering an area of about \SI{500}{\kilo\metre\squared}, complementing the optical part of the experiment \cite{TDR}. Currently, two detector components are planned to detect neutrinos with the IceCube-Gen2 Radio array. A 'shallow' station, with antennas close to the surface and a 'hybrid' station, itself consisting of a 'shallow' component and a 'deep' component with antennas down to \SI{-150}{\meter}. While located in close proximity to each other, the 'shallow' and 'deep' components of a 'hybrid' station would operate with independent triggers, allowing us to treat them separately. 

The Radio Neutrino Observatory (RNO-G), currently under construction in Greenland, uses a similar approach by deploying 35 'hybrid' stations \cite{Aguilar_2021}. Although this work focuses on reconstructing events detected by the IceCube-Gen2 Radio detector in South Polar ice, the developed neural network is highly modular and can easily be adjusted to the different detector geometry of RNO-G and the Greenlandic ice sheet.

The reconstruction presented in this work utilizes a large neural network (initial convolutional encoding followed by several convolutional blocks with residual connections), which simultaneously predicts the posterior PDF of the neutrino direction and the shower energy of the neutrino-induced interaction with a significantly better resolution compared to previous analyses. Furthermore, for the first time, we predict the full posterior PDF of the estimated energy and direction, allowing us to quantify event-by-event uncertainties. We achieve this by combining neural networks with conditional normalizing flows \cite{norm_flow}. This is particularly useful for highly non-Gaussian uncertainty contours, which are often encountered for this type of detector \cite{dir_reco}. In addition, the model returns a percentage of how sure it is that the event came from an electron neutrino $\nu_e$ - CC interaction with an electromagnetic shower component or from a NC interaction with a single hadronic shower. This method requires no prior knowledge about the event topology, and no analysis cuts are applied to the simulated data. The model also predicts the vertex coordinates (x, y, and z) of the interaction and their correlation to each other and the predicted shower energy. Together, all of the predictions uniquely define the in-ice shower and allow to calculate the expected radio emission for the reconstructed parameters. This enables us to construct a goodness-of-fit score between the (noisy) measured data and the reconstructed neutrino signal, which can be used as an experimental verification that the neural network describes the measured data correctly, an important advancement given that the network was only trained on simulated data. We perform the analysis for both of the proposed detector components ('shallow' and 'deep') by training two separate models. The model architectures for both are identical, except for minor differences in the hyperparameter settings tuned for each detector component. Furthermore, we were able to extract information about the impact of different antenna types in regards to the resolution, which can be used to inform future detector designs. Additionally, the systematic uncertainties for the ice model and the antenna position/orientation provide insight into the required accuracy for firn measurements of the refractive index and the accuracy of antenna deployment.

For this paper, we will discuss the considerations when reconstructing in-ice radio neutrino events in section \ref{sec: radio} and lay out the specifications of the Monte Carlo data generation in section \ref{sec: Monte}. In section \ref{sec: Network}, we walk through the neural network. The results for the energy, direction, and flavor reconstruction are presented in the sections \ref{sec: Results_energy}, \ref{sec: Results_direction}, and \ref{sec: Results_flavor}, respectively, together with the impact of different antenna types. The systematic uncertainties are discussed in section \ref{sec: Systematic}, and we lay out the construction for a goodness-of-fit score of the reconstructed neutrino signal in section  \ref{sec: Goodness}. The conclusion of the analysis can be found in section \ref{sec: Conclusion}.

\section{In-ice Radio Reconstruction}\label{sec: radio}

In the following, we briefly describe how the main neutrino parameters of interest (energy, direction, and flavor) impact the timing, shape, and amplitude of the emitted and observable radio signal \cite{Barwick:2022vqt}, to provide an intuition for interpreting the resolution of the neural network. 

The neutrino energy is proportional to the measured signal amplitude. However, the signal is attenuated as it propagates through the ice, reducing its amplitude, creating an inverse correlation between the distance to the interaction vertex and the neutrino energy. This makes the vertex position the biggest challenge in the shower energy reconstruction. Furthermore, the inelasticity of the neutrino interaction, the loss of coherence when emitted slightly off of the Cherenkov cone (viewing angle), the polarization of the electric field, and the sensitivity of the antenna have to be accounted for. As the inelasticity in neutral current (NC) interactions is a random property, there is no possibility of directly inferring the neutrino energy from the reconstructed shower energy. Therefore, the neutrino energy resolution will always be limited by an unavoidable uncertainty of $\sim$0.3 in log(E) \cite{Anker_2019} in addition to the statistical uncertainties presented in this work. However, as we developed a method of differentiating $\nu_x$ - NC and $\nu_e$ - CC , this limit does not apply to events where the reconstruction is very certain that the event came from a CC interaction (in this case, the reconstructed shower energy would be equivalent to the neutrino energy). 

The neutrino direction is not the direction from which the radio signal arrives at the antennas. Signals emitted on the Cherenkov cone have a launch angle of $\sim$\SI{56}{\degree} with respect to the neutrino direction, and they quickly lose coherence as the launch angle moves away from the Cherenkov cone \cite{nuradiomc}. Also, while propagating through the ice, the signal trajectories are bent downwards in the upper $\sim$\SI{200}{\meter} of the ice sheet due to a pressure gradient and the resulting change in refractive index \cite{Barwick_2018}. Finally, the polarization vector of the emitted signal always points towards the shower axis, constraining the location on the Cherenkov ring. The polarization is of particular interest here, as previous analyses have shown that its reconstruction is the biggest challenge when reconstructing the neutrino direction \cite{dir_reco}. This difficulty leads to non-Gaussian uncertainty shapes, as the uncertainty contour of the neutrino direction is a segment of the Cherenkov ring projected on the sky. As the polarization reconstruction gets better, the ring segment gets smaller and the uncertainty contour approaches a more Gaussian shape. Previous reconstructions often quoted their results in terms of space-angle-difference, a single angle between the true and reconstructed direction vector. Due to the difficulty in polarization reconstruction and the resulting non-Gaussian uncertainty contours, the better method is to quote the area of the uncertainty contour size in square degrees.

With three neutrino flavors and neutral-current ($\nu_x$ - NC) and charged-current ($\nu_x$ - CC) interactions, there are in principle six event topologies to consider (twelve when including anti-neutrinos). However, the $\nu_x$ - NC interactions for all flavors produce a single hadronic shower, leading to the same signal in the detector. Furthermore, the CC interactions of tau and muon neutrinos also produce a hadronic shower and a lepton that escapes the detector volume of a single radio station, again leading to the same signal as the $\nu_x$ - NC interactions in the detector. Only the charged current electron neutrino interactions ($\nu_e$ - CC) induce a different signal as the produced electron deposits its energy in one or several electromagnetic showers very close to the original neutrino interaction. Therefore, only two event topologies are considered here: $\nu_x$ - NC (representing $\nu_e$-NC, $\nu_\mu$-NC, $\nu_\tau$-NC, $\nu_\mu$-CC, and $\nu_\tau$-CC) and $\nu_e$ - CC (only representing $\nu_e$-CC). Furthermore, the highly energetic electrons created in $\nu_e$ - CC interactions are impacted by the LPM-effect \cite{Landau:1953um,PhysRev.103.1811}, changing their cross section. This leads to stochastically different particle shower profiles, which alter the emitted radio signals \cite{lpm_neutrino,nuradiomc}, making them more difficult to reconstruct. However, this effect can also be used to identify electron charged-current interactions and thereby give insight into the flavor composition of UHE neutrinos at the detector and at their sources \cite{PhysRevD.110.023044}

So far, no neutrino has been detected with the radio technique, and current or future detectors project a sensitivity to only a few annual events for optimistic flux scenarios. Therefore, to study the reconstruction capabilities, Monte Carlo simulations have to be performed, generating a representative data sample to which reconstruction algorithms are applied. So far, the best reconstruction performance of neutrino properties from in-ice emitted radio signals was reached with the forward-folding technique \cite{NuRadioReco}. Here, the shower (and thereby neutrino) properties are adjusted so that the resulting radio signal, propagated through the ice and folded with the detector response, gives the best match with the measurement. With this technique, a direction resolution of $\Delta \Psi_{68\%} = 2.9^{\circ}$ was achieved for $\nu_x$ - NC events at a shower energy of \SI{1}{\exa\eV} for the ARIANNA experiment which comprised 'shallow' detector stations equivalent to the 'shallow' detector components presented here \cite{ARIANNADirectionICRC2021}. For the 'deep' component of the RNO-G detector, an energy resolution of 30\% was achieved for $\nu_e$ - NC events after moderate analysis cuts \cite{energy_reco}. For the 'deep' component of the IceCube-Gen2 Radio detector, a 68th percentile reconstruction error of $\sim\SI{1000}{}$ square degrees (corresponding to a symmetric 1D uncertainty of $\sim\SI{11}{}$ degrees) was achieved \cite{dir_reco_POS}. The 'deep' component of the RNO-G detector achieved similar results \cite{dir_reco}. However, this technique does currently not predict event-by-event uncertainty contours, making a direct comparison difficult. Also, it is difficult to apply the forward-folding technique to $\nu_e$ - CC events due to the stochastic variation of the shower profile from the LPM effect, and the forward-folding requires a deterministic model of the shower profile. 

In recent years, machine learning and especially deep learning have advanced significantly. In this approach, all the 'physics' knowledge of the analysis is in the simulated dataset. After that, the training of the neural network itself is performed in a supervised manner, comparing the reconstructed properties to the true parameters from the Monte Carlo simulation and optimizing the weights of the model via back-propagation until the model converges. This method has been shown to significantly improve the scientific output of experiments due to the large number of tunable parameters capable of exploiting minor features in the data \cite{doi:10.1126/science.adc9818,Auger_Xmax,Auger_Masscomp}. However, this method is reliant on the accuracy of the Monte Carlo dataset used to train the models, as discrepancies can negatively impact the reconstruction results. One previous study used a deep learning approach to estimate the neutrino energy and direction for 'shallow' antennas of the IceCube-Gen2 Radio detector \cite{GLASER2023102781}. The model achieved a shower energy resolution of 0.3 in log(E) for both $\nu_x$ - NC and $\nu_e$ - CC events (however, the reconstruction had a strong energy-dependent bias of 0.4 to -0.2 in log(E) between $\SI[parse-numbers=false]{10^{17}}{\eV}$ - $\SI[parse-numbers=false]{10^{19}}{\eV}$ of neutrino energy). The direction reconstruction yielded a resolution of $\Delta \Psi_{68\%} = 4^{\circ}$ for $\nu_x$ - NC events and $\Delta \Psi_{68\%} = 5^{\circ}$ for $\nu_e$ - CC events. Another study used deep learning to estimate the vertex position and the neutrino direction of 'deep' antennas with a resolution of $4^{\circ}$ in the zenith angle and $6^{\circ}$ in the azimuth angle of the neutrino arrival direction \cite{ARA_reco}. However, this reconstruction used high-level features (maximum signal amplitude, signal time) instead of the voltage traces, and included timing information from the Monte Carlo labels during training.

\section{Dataset Generation}\label{sec: Monte}

The data used to train and test our neural network was generated with the Monte Carlo framework NuRadioMC and NuRadioReco \cite{nuradiomc,NuRadioReco}. Specifically developed for in-ice radio neutrino detection, NuRadioMC and NuRadioReco are capable of simulating every aspect of the detection process, starting from the initial neutrino interaction, the radio emission from the induced particle shower, the propagation of the radio signals through the ice, up to the antenna response of the simulated detector. This allows us to simulate many millions of neutrinos and store the signals that triggered the stations. In this way, we record the signals a neutrino would produce in our antennas while also retaining the Monte Carlo truth of the neutrino properties. This sets up the reconstruction as a supervised deep learning problem. For each of the two station components, we simulated 2.1 million neutrino interactions that fulfill the trigger condition, with 1.8 million used for training, 0.2 million used for validation, and 0.1 million used only to produce the final results.

\subsection{Detector Layout}\label{sec: Layout}

Here, we present the considerations that were made about the detector layout when simulating the events. Although the full detector of IceCube-Gen2 radio is planned to be an array of more than 350 stations, they operate mostly independently of each other so that a single neutrino interaction will only be seen by more than one station in $\sim$10\% of the cases \cite{TDR}. For that reason, this analysis focuses on the reconstruction capabilities of a single station without considering potential coincidences. 

\begin{figure}[tbp]
  \centering
  \includegraphics[height=4in]{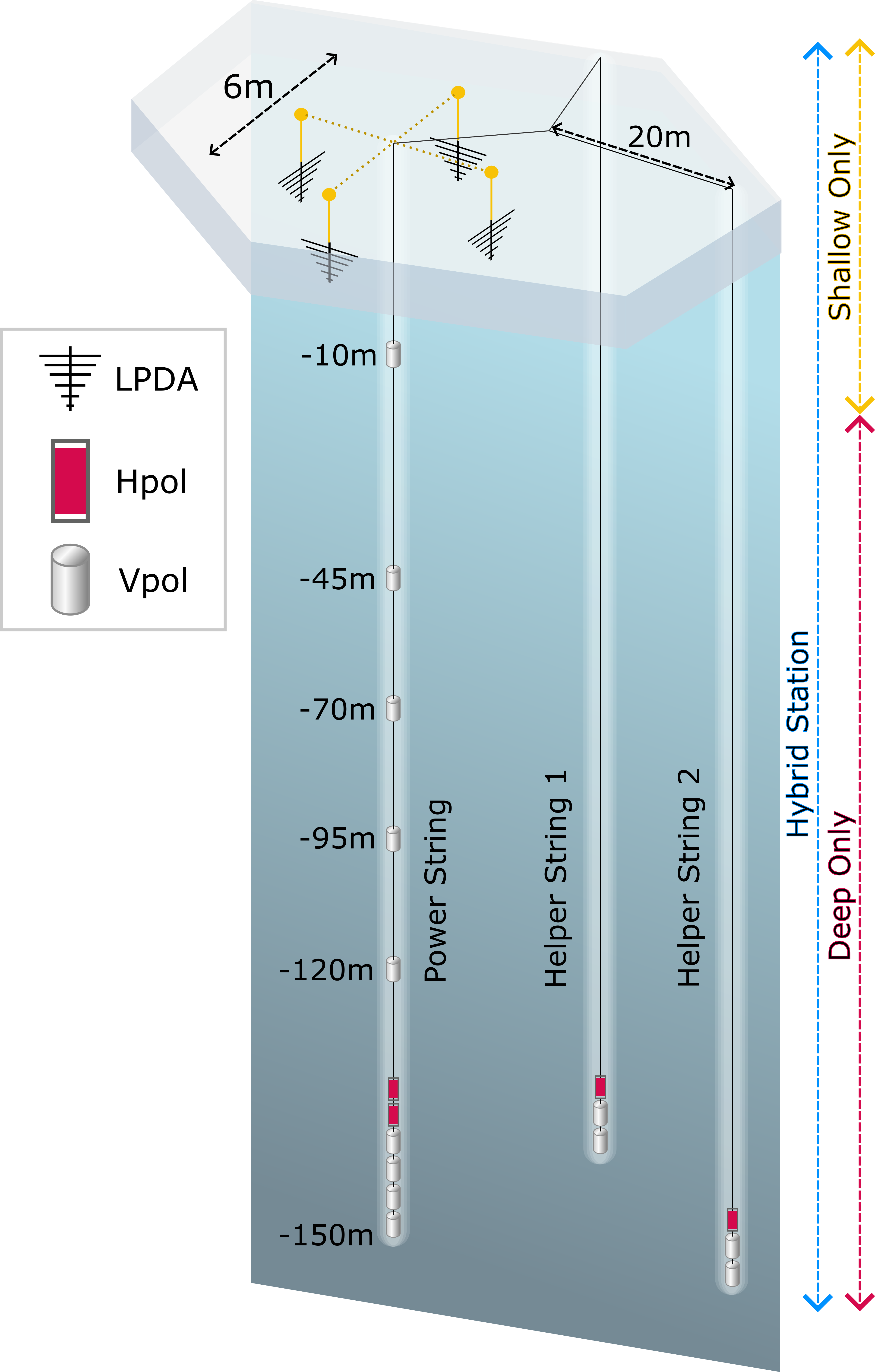}
  \caption{'Hybrid' radio station with the antennas relevant for neutrino detection as envisioned for IceCube-Gen2 Radio. The 'shallow' component of the station encompasses four LPDA antennas and one Vpol antenna, while the 'deep' component of the station encompasses 12 Vpol antennas and four Hpol antennas.}
  \label{fig:detector}
\end{figure}

There are two station components planned to be deployed for IceCube-Gen2 Radio \cite{TDR}. The 'shallow' detector component (see figure \ref{fig:detector}, top) consists of four downward-pointing LPDA antennas deployed a few meters below the snow surface and a single bicone Vpol antenna at a depth of \SI{-10}{\meter}. Additional upward-facing antennas that can be used to detect cosmic rays were ignored for this analysis. To be able to see differently polarized pulses, the four downward-facing antennas are planned to be \SI{6}{\meter} apart on a horizontal square. The trigger for the 'shallow' antennas consists of a 2 out of 4 coincidence trigger \cite{Steven_Barwick_2007} of the 4 downward-facing LPDA antennas with a high-low threshold trigger where the threshold corresponds to a \SI{100}{\hertz} trigger rate on thermal noise. As the LPDA antennas are all within \SI{6}{\meter} of each other and the central bicone Vpol antenna is only \SI{-10}{\meter} in the ice, the neutrino-induced radio signals will be relatively close to each other in time. At a sampling rate of \SI{2.4}{\giga \hertz}, most signals will be visible inside a window of 512 samples corresponding to \SI{213}{ns}. Therefore, each event will produce an array of shape (5, 512).

The 'hybrid' station design (see figure \ref{fig:detector}) consists of all the antennas of the 'shallow' design as well as 16 additional 'deep' antennas deployed in three boreholes reaching down to \SI{-150}{\meter}. The 'power string' deployed in the center of the 'shallow' antennas holds (apart from the 'shallow' bicone Vpol antenna at \SI{-10}{\meter}) four bicone Vpol antennas at \SI{-45}{\meter}, \SI{-70}{\meter}, \SI{-95}{\meter}, and \SI{-120}{\meter}, as well as two slotted cylinder Hpol antennas at \SI{-141}{\meter}, and \SI{-142}{\meter} and four more bicone Vpol antennas at \SI{-147}{\meter}, \SI{-148}{\meter}, \SI{-149}{\meter}, and \SI{-150}{\meter}. The two helper strings both hold a single slotted cylinder Hpol antenna at \SI{-142}{\meter} and two bicone Vpol antennas at \SI{-143}{\meter}, and \SI{-144}{\meter}. The strings in the three boreholes are ~\SI{35}{\meter} apart horizontally on an equilateral triangle. The trigger for the 'deep' antennas consists of a four-antenna phased-array trigger \cite{ALLISON2019112} of the four deepest bicone Vpol antennas on the power string, where the threshold corresponds to a \SI{100}{\hertz} trigger rate on thermal noise. As the 'deep' antennas are spread over ~\SI{100}{\meter} in spatial separation, the neutrino-induced radio signals will be relatively far from each other in time. At a sampling rate of \SI{2.4}{\giga \hertz}, most signals will be visible inside a window of 2046 samples corresponding to \SI{853}{ns}. Therefore, each event will produce an array of shape (16, 2046).

While deployed within a single station, the 'shallow' and 'deep' components of a 'hybrid' station operate independently of each other, with different triggers used for each of them. Therefore, the event reconstruction was performed independently for the five 'shallow' antennas and the 16 'deep' antennas. It is worth mentioning that neglecting coincidences between different station components and between multiple stations are conservative assumptions, as some events are expected to be seen in both components of a hybrid station or even in multiple stations, which would significantly improve the reconstruction as more information would be available for a dedicated reconstruction algorithm. The antenna response from the three described antenna types used for this study (LPDA, Vpol, Hpol) follows the specifications in the IceCube-Gen2 technical design report and previous reconstruction studies \cite{TDR,dir_reco}.

\subsection{Simulation Specifications}

Apart from the above-mentioned differences in the detector layout and trigger, the two datasets created for the 'shallow' and 'deep' antennas follow the same simulation specifications. Even though the distribution of expected events consists of different fractions of $\nu_x$ - NC and $\nu_e$ - CC events, the datasets used for this analysis contain the same amount of each in order to avoid a bias when trying to distinguish between the two.

We simulate a uniform spectrum of neutrino arrival directions as the neutrino flux is expected to be isotropic. However, the Earth is mostly opaque to neutrinos at these energies. Therefore, each event is given a probability of reaching the detector volume. If the probability is below 0.0001\%, the event is cut from the dataset. However, this also means that two events, both passing the threshold, but with different weights, are used equally when training the models. In this way, events coming from the region below the horizon are over represented in the dataset. For many science analyses, such as the neutrino cross-section measurement, the events from these regions are particularly interesting, which means it is good that these events are both strongly represented in the dataset and unbiased with regards to the zenith spectrum. Furthermore, most neutrinos arriving with a steep zenith angle from above the detector are unlikely to trigger the stations. For these reasons, the events that fulfill the trigger condition, i.e., the events that end up in the training data set, will still have a uniform azimuth distribution while the zenith spectrum peaks for moderate angles from above the horizon.

For the energy reconstruction, we decided to reconstruct the shower energy instead of the neutrino energy, as the energy transfer in a neutral-current interaction is fully stochastic and could only be estimated statistically. Here, shower energy refers to all the energy that is deposited into particle cascades in the ice. This is the energy from a single hadronic shower in case of a $\nu_x$ - NC event, or the neutrino energy in case of a $\nu_e$ - CC interaction. A realistic shower energy spectrum can be calculated by folding the expected neutrino flux with the detector sensitivity. However, previous deep learning based analyses have shown an energy-dependent bias in the energy reconstruction if the high- and low-ends of the energy spectrum contain a low number of events \cite{GLASER2023102781}. For that reason, we decided to force the shower energy spectrum of triggered events to be uniform, reaching from $\SI[parse-numbers=false]{10^{16.0}}{\eV}$ up to $\SI[parse-numbers=false]{10^{20.2}}{\eV}$. This uniform spectrum helps to keep the network predictions unbiased over the relevant energy range $\sim\SI[parse-numbers=false]{10^{17.0}}{\eV}$ to $\sim\SI[parse-numbers=false]{10^{19.0}}{\eV}$. All results are shown as a function of shower energy, such that they can be applied to an arbitrary spectrum. 

We use the ARZ Askaryan emission model \cite{PhysRevD.101.083005} for the signal generation, model the ice with an exponential ice profile \cite{Barwick_2018}, and apply the detector temperature of \SI{300}{\kelvin} to generate thermal noise. Apart from thermal noise, no other potential noise classes, such as anthropogenic noise, were considered when training the models. About 9\% of mostly high SNR events were removed before training from the 'shallow' dataset due to nonphysical simulation artifacts. However, this did not bias the training in a significant way due to the large amount of events remaining in the dataset.

\section{Network Architecture}\label{sec: Network}

\begin{figure*}[tbp]
  \centering
  \includegraphics[trim=0cm 3cm 0cm 2.5cm, clip, width=5in]{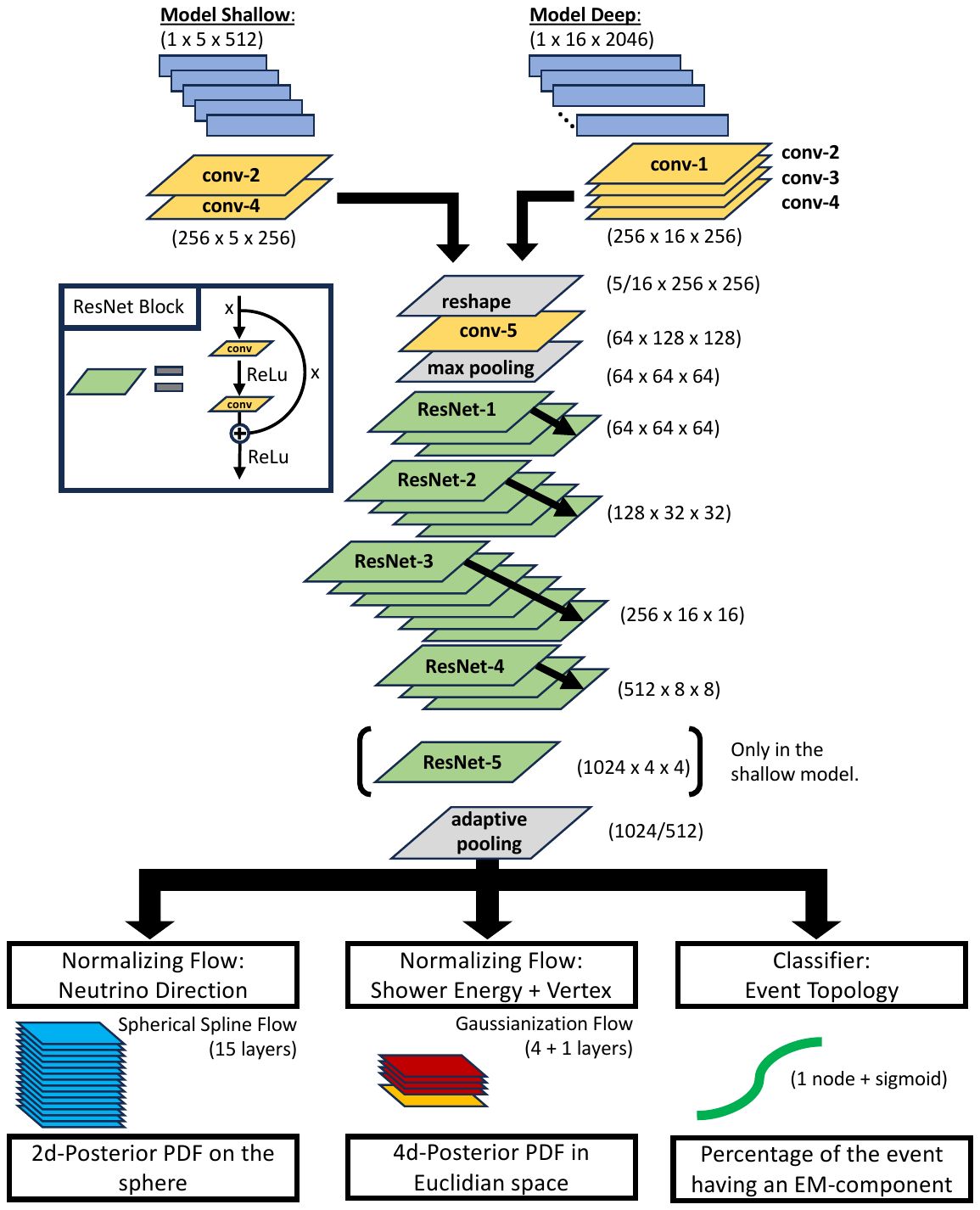}
  \caption{Neural network architecture schematic developed for this reconstruction. Starting from the input data at the top, the data flows through several 1d convolutional layers before being handed to multiple blocks of ResNet layers. After pooling the output from the ResNet Blocks the compressed data is handed to the conditional normalizing flows for the direction and energy prediction, as well as a classifier for the event topology prediction. Two separate models were trained for the 'shallow' and 'deep' detector components but large parts of the network architecture are identical, with differences arising during hyperparameter tuning.}
  \label{fig:architecture}
\end{figure*}

For our neural network, we combine several architectures into one large model. A schematic of the data flow through the model can be seen in figure \ref{fig:architecture}. The input data are the antenna waveforms with dimensions (1 x 5 x 512) for the 'shallow' component and (1 x 16 x 2046) for the 'deep' component. Here, the '1' corresponds to the feature dimension, i.e., the measured amplitude in the case of the input layer, '5' or '16' corresponds to the number of antennas, and '512' or '2046' corresponds to the number of time-samples in each trace. In the first section of the architecture, we use several one-dimensional convolutional layers to downsample the time dimension while up-sampling the feature dimension. For the 'shallow' detector component, we use two blocks with four 1d convolutional layers each. The first block uses 64 filters with a kernel size of 16, after which an average pooling layer with a kernel size of 2 is applied. The second block uses 256 filters with a kernel size of 16. All convolution layers share the weights over the antennas. For the 'deep' detector component, we use four blocks with four 1d convolutional layers each. The first block uses 32 filters, the second 64 filters, the third 128, and the fourth 256, each with a kernel size of 16 and an average pooling layer with a kernel size of 2 at the end of the block (except the last one). This first part of the architecture was inspired by the success of the previous deep learning analysis for radio detectors \cite{GLASER2023102781}. After the one-dimensional blocks, the data arrays are reshaped into (5 x 256 x 256) and (16 x 256 x 256), where the first dimension still corresponds to the antenna dimension. After this step, the event is treated as an image of the size (256 x 256) pixels, but instead of three color channels, we use 5 or 16 'antenna' channels. 

In the second section of the model, a ResNet architecture \cite{He2015DeepRL} (modified to handle more than three color channels) is applied to these 'images'. This is where the bulk reconstruction capability of the model comes from, and the approach was inspired by a Kaggle challenge on reconstructing gravitational waves \cite{grav_waves}. The modified ResNet architecture connects the output of convolutional layers with residual connections from previous layers, thereby allowing for very deep networks without vanishing gradients. Several other options for this section were also investigated, such as conventional convolutional layers or larger ResNet architectures with the presented architecture yielding the best results. For the 'shallow' model, an additional ResNet block was added at the end due to a slightly better performance (The same tests were made for the 'deep' model, where the performance did not increase). The output of the ResNet structure is compressed into 1024/512 nodes for the 'shallow'/'deep' model via adaptive pooling (a concatenation of max-pooling and average-pooling). 

Then these 1024/512 nodes are handed to three independent output structures (two conditional normalizing flows which predict the posterior direction and energy distribution, and a binary classifier for the event topology). The conditional normalizing flows were integrated into the models using the jammy flows library \cite{thorsten_jammy,jammy}, allowing for easy implementation and stable convergence. In the first output structure, 15 spherical spline flows model a PDF on the two-dimensional sphere to predict the neutrino direction. In the second output structure, 4 gaussianization flows \cite{meng2020gaussianization} plus an additional multivariate normal flow model a PDF in four-dimensional Euclidean space to predict the shower energy and vertex position (x, y, z), including their correlations. Correlations between the direction and the energy are ignored in this approach. However, several other flow options were explored, with the one presented here yielding the best results. In the classifier, the 1024/512 nodes are connected to a single output node with sigmoid activation such that the output of this node can be interpreted as a percentage of the model's certainty that the event contained an electromagnetic component. The loss of the model used in the backpropagation is a sum of the losses of the three output structures, where the conditional normalizing flows simply use the negative log-likelihood of the predicted PDF at the point of the MC-true label, and the classifier uses the binary cross-entropy between the predicted and the MC-true event topology. The weights by which these loss terms are added are hyperparameters tuned for the two models.

The training was performed with an Adam optimizer and a scheduler, which reduced the learning rate when the validation loss plateaued. Early stopping was performed when the model did not improve after 10 epochs, and the model with the lowest validation loss was chosen to generate the obtained resolution. All trainings relevant to this work were performed on datasets where $\nu_x$ - NC and $\nu_e$ - CC events, as well as all shower energy bins, were randomly shuffled across the training process to avoid overfitting to the training data. It is important to note that the architecture is set up in a modular way, making it very easy to adapt to different input dimensions should the number of antennas or samples per antenna change in the future (or if this method is applied to different detectors such as RNO-G).

\section{Neural Network Performance}\label{sec: Results}

\begin{figure*}[tbp]
  \centering
  \includegraphics[height=3.5in]{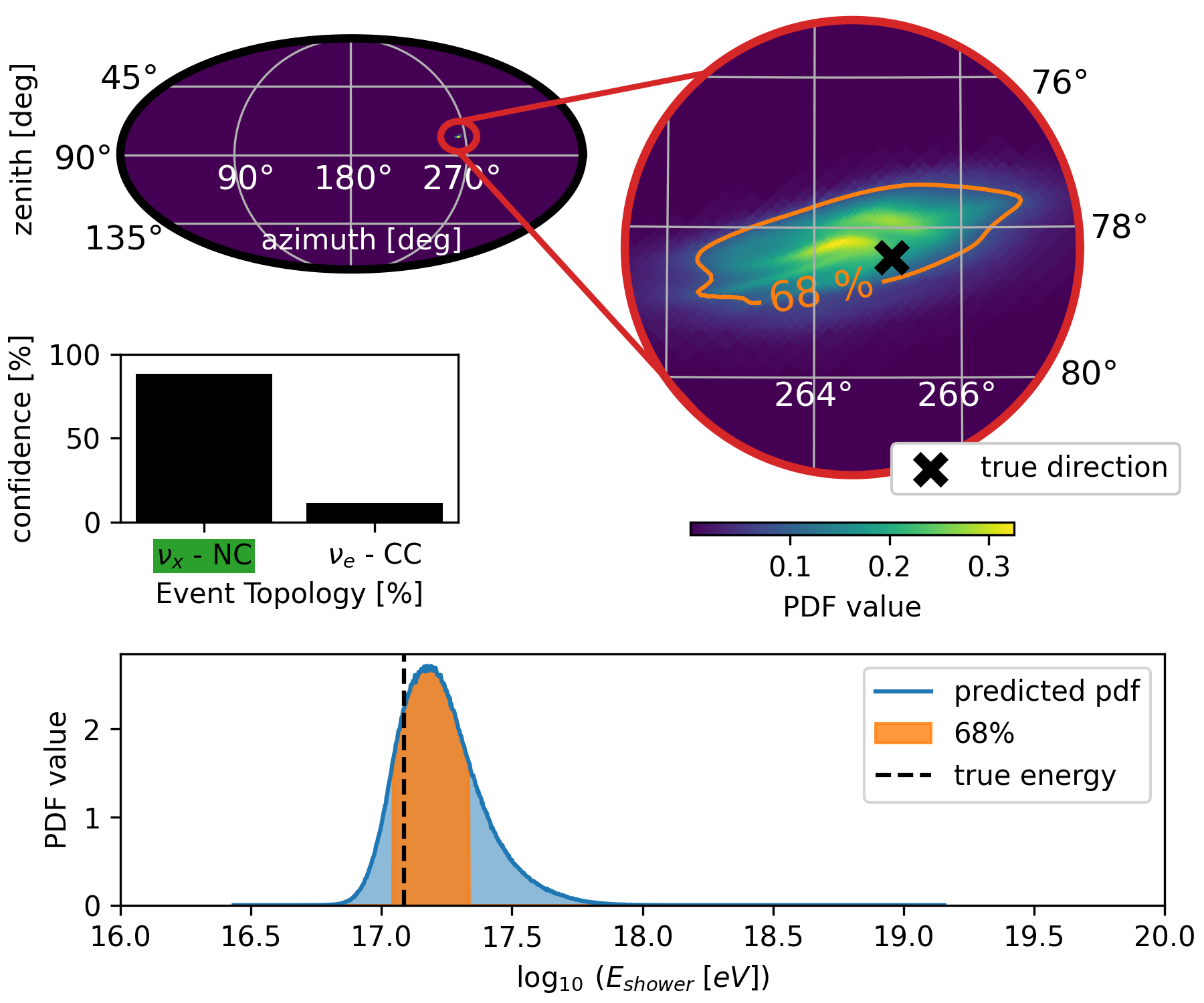}
  \caption{Overview of the model output for a single event. Top left: Full sky map with the predicted neutrino direction PDF visible as a small dot. Top right: Zoomed-in sky map of the PDF of the predicted neutrino direction. The 68\% uncertainty contour of the predicted PDF is shown in orange, and the MC true direction of the event is shown as a black cross. Center-left: Classification results where the model predicts a confidence for the '$\nu_x$ - NC' or '$\nu_e$ - CC' event topology, while '$\nu_x$ - NC' was the MC true event topology. Bottom: Predicted shower energy PDF with the 68\% uncertainty region indicated in orange, and the MC true shower energy indicated as a dashed line. This is the marginalized distribution from the 4-dimensional PDF, which also includes the vertex position. The correlations between the shower energy prediction and the vertex position prediction for the same event can be seen in figure \ref{fig:correlations}, and the event trace can be seen in figure \ref{fig:shallow_trace}.}
  \label{fig:overview}
\end{figure*}

When applying the developed neural network to a single test event, we can produce an overview of all relevant properties of the reconstruction. An example of this can be seen in figure \ref{fig:overview}. It shows the predicted posterior PDF for the shower energy and neutrino direction and compares it to the Monte Carlo true parameters with which the data was produced. It also displays the confidence in percent that the event was produced by the $\nu_x$ - NC or $\nu_e$ - CC event topology. The predicted vertex position and its correlation to the predicted energy are not shown but discussed in \ref{sec: Correlation}. 

To gain insight into the resolution of the reconstruction, we evaluate the full test dataset. We first check the accuracy of the predicted uncertainties by calculating the coverage, i.e., for every test event, we checked in which percentile of the predicted PDF the MC-true value lies. The results are shown in \ref{sec: Coverage}. We find some under-coverage between 8\% and 17\% (meaning the predicted PDFs were slightly too small). To still be able to correctly quantify the reconstruction resolution, we applied an energy-dependent correction. The correction was calculated by first checking the coverage in a certain energy bin and then assessing the size of the contour in the corrected percentile. For example, if we want to calculate the size of the 68\% uncertainty contour but have 5\% under-coverage, we measure the size of the 73\% uncertainty contour instead to ensure that 68\% of the MC-true events lie in this contour.

\subsection{Energy Resolution}\label{sec: Results_energy}

\begin{figure*}[tbp]
  \centering
  \includegraphics[height=2.5in]{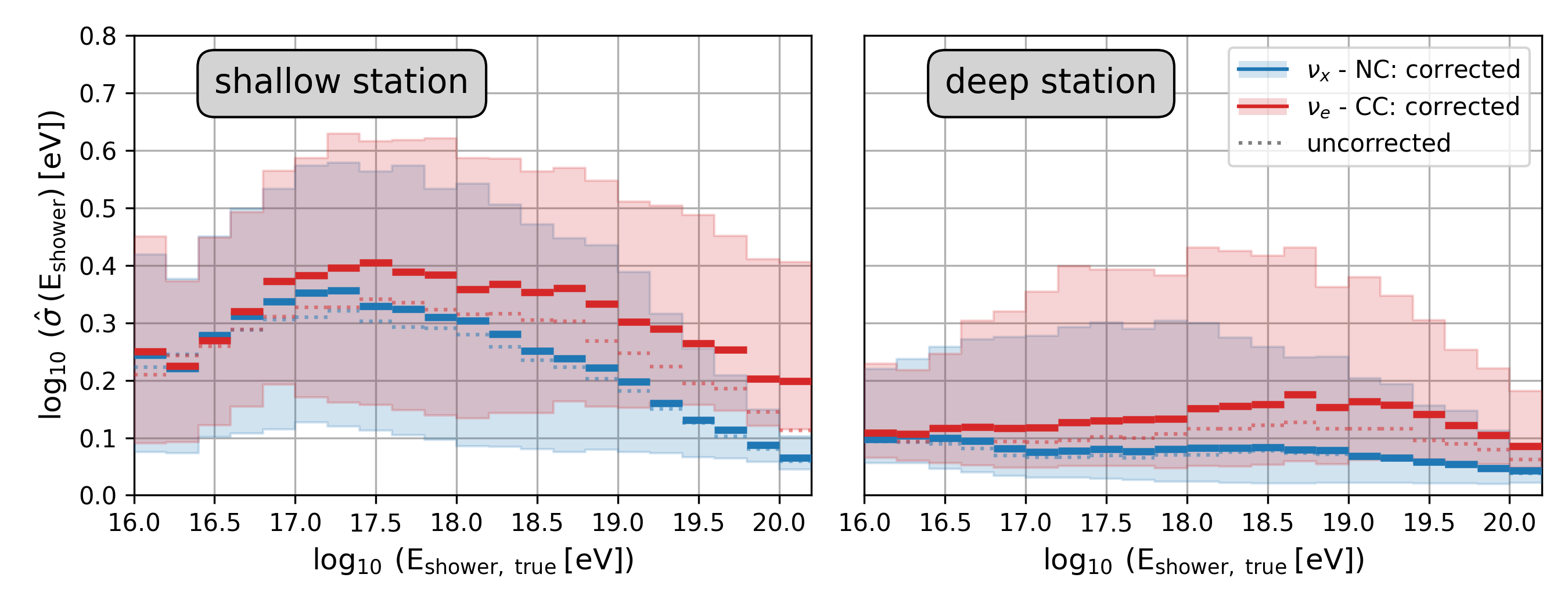}
  \caption{Results for the shower energy reconstruction on the test dataset. The median of the uncorrected resolution is indicated as a dotted line, while the median for the corrected resolution per energy bin is shown as a solid line. The shaded areas indicate the 16th and 84th percentiles of the resolution per bin. The results for the '$\nu_x$ - NC' topology are blue while the results for the '$\nu_e$ - CC' topology are red. Left: Results for the 'shallow' station component. Right: Results for the 'deep' station component.}
  \label{fig:energy_resolution}
\end{figure*}

To determine the shower energy of an event, the network uses a 4-dimensional Euclidean normalizing flow that predicts the shower energy of the event, as well as the coordinates of the vertex position. By including the vertex position as an additional constraint in the reconstruction, we observed improved performance in determining the shower energy. It also allowed us to construct the goodness-of-fit score in section \ref{sec: Goodness}, where the location of the interaction is crucial for the recreation of the event. However, the main parameter of interest is the shower energy, which we present in the following.

Figure \ref{fig:energy_resolution} shows the shower energy resolution on the test dataset as a function of the shower energy. The shower energy resolution is measured by calculating the half-width of the 68\% highest density interval (HDI). This way, the resolution can be calculated for non-Gaussian, multi-modal distributions, but it converges to the standard deviation as the predicted PDF approaches a Gaussian. 

\begin{figure}[tbp]
  \centering
  \includegraphics[height=5in]{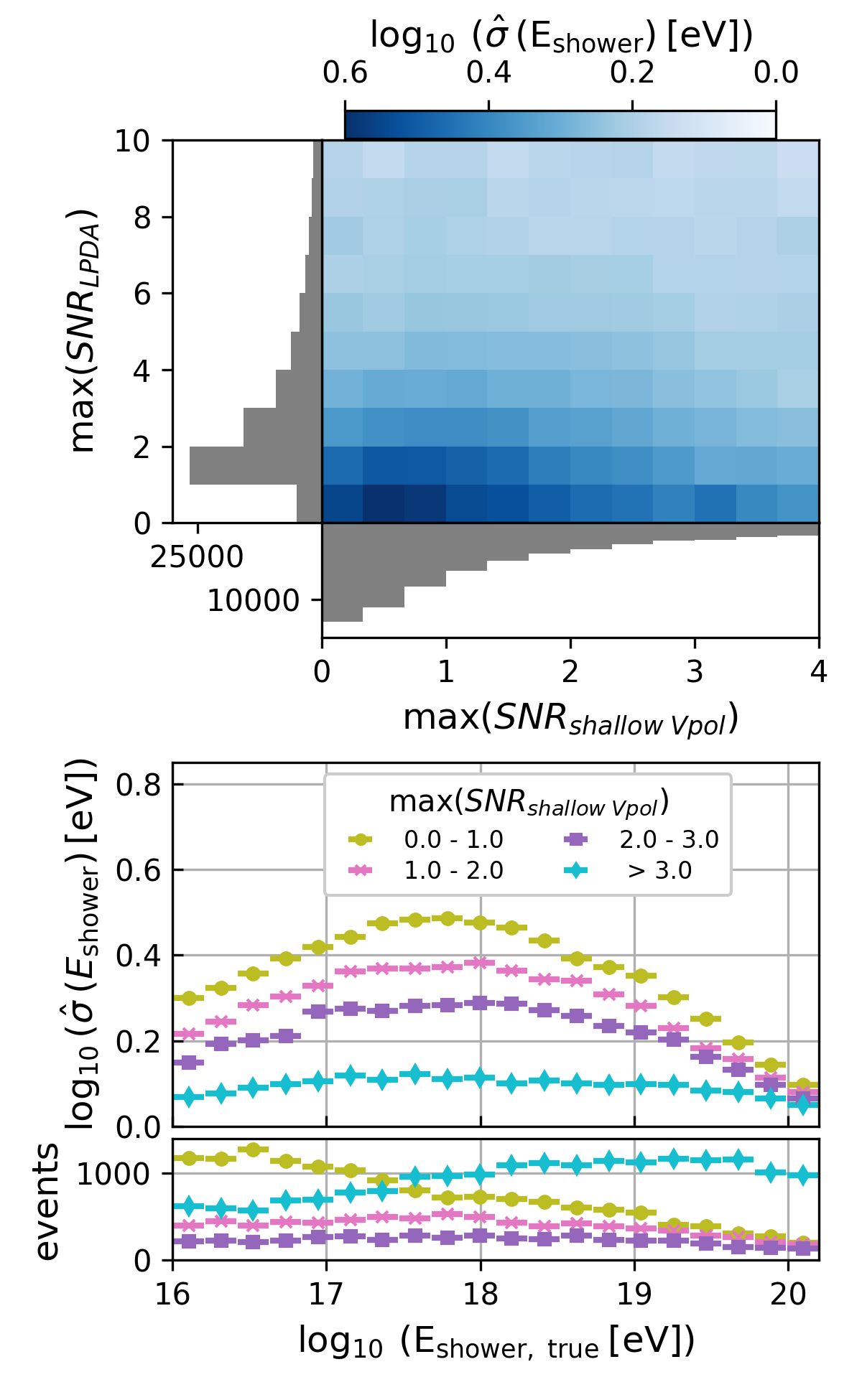}
  \caption{Impact of the 'shallow' antenna SNR on the shower energy resolution for '$\nu_x$ - NC' events. Top: The maximum SNR of any of the ray tracing solutions in any of the 4 LPDA antennas plotted against the maximum SNR of any of the ray tracing solutions for the Vpol antenna. The color scale indicates the median shower energy resolution per bin, where a lighter shade of blue means a better resolution. The gray histograms indicate how many events are in each of the bins. However, this is not a physical SNR distribution as the data was generated in uniform energy bins. Bottom: The corrected median shower energy resolution per shower energy bin for different Vpol SNR bins. The number of events is shown in the plot below.}
  \label{fig:shallow_SNRvsE}
\end{figure}

For the 'shallow' station component, we obtain a median 68\% uncertainty between 0.2 and 0.35 in log(E) for $\nu_x$ - NC events and between 0.3 and 0.4 in log(E) for $\nu_e$ - CC events. Below $\SI[parse-numbers=false]{10^{17}}{\eV}$, the uncertainty increases with shower energy while above $\SI[parse-numbers=false]{10^{17}}{\eV}$ it decreases. We attribute this effect to the inverse correlation between shower energy and vertex distance. The amplitude of the radio signal arriving at the antenna is proportional to the energy but inversely proportional to the vertex distance. As we only use signals that passed the trigger, at low shower energies, the dataset includes mostly events originating at a close proximity to the station. These events have a strongly bent wavefront, making their vertex position and, therefore, shower energy reconstruction easier, improving the obtained resolution. Furthermore, these events usually only pass the trigger when emitted very close to the Cherenkov angle, also resulting in an easier reconstruction. At high shower energies, the dataset includes events that can have originated several kilometers away. However, the detector volume is vertically limited to the $\sim$\SI{3}{\kilo\meter} thick ice sheet such that for these events a higher signal amplitude usually corresponds to a higher shower energy, again making these events easier to reconstruct. These effects can be disentangled when including the antenna SNR, where an increase in SNR of the triggering antenna always corresponds to a better resolution (seen in figure \ref{fig:shallow_SNRvsE}). 

Although the model is agnostic to the event topology, it learned that the shower energy of $\nu_x$ - NC events can be determined more precisely than for $\nu_e$ - CC events. The discrepancy in resolution between the two increases with shower energy as the effects of the LPM effect become more pronounced and the signal for $\nu_e$ - CC events becomes more stochastically varied. However, the resolution of the neutrino energy will increase significantly for $\nu_x$ - NC events due to the inelasticity of the interaction, while the resolution for $\nu_e$ - CC events will remain the same for shower and neutrino energy as long as the event was classified correctly. When reconstructing $\nu_e$ - CC events, the model struggled more with under-coverage compared to $\nu_x$ - NC events, especially at high energies (see figure \ref{fig:comb_cov}, 4\% for $\nu_x$ - NC and 9\% for $\nu_e$ - CC events).

For the 'deep' station component, we obtain a median 68\% uncertainty between 0.05 and 0.1 in log(E) for $\nu_x$ - NC events and between 0.1 and 0.2 in log(E) for $\nu_e$ - CC events. Here, we observe a more subtle behavior of the resolution as a function of shower energy, but it is still visible, especially in the 84th percentile, that it is not a strictly falling relationship. Also, we observe better performance for $\nu_x$-NC events than for $\nu_e$-CC events, with the discrepancy increasing with energy because the LPM effect becomes more pronounced, resulting in a more stochastic development of the $\nu_e$-CC initiated showers. Furthermore, when reconstructing $\nu_e$ - CC events, the model struggled more with under-coverage compared to $\nu_x$ - NC events, especially at high energies (see figure \ref{fig:comb_cov}, 5\% for $\nu_x$ - NC and 13\% for $\nu_e$ - CC events).

For the 'shallow' station component, a shower energy resolution of 0.3 in log(E) has been reported in a previous reconstruction study \cite{GLASER2023102781} for both $\nu_x$ - NC and $\nu_e$ - CC events across the relevant energy range. This resolution is similar to the results we found with our neural network. However, the previous analysis showed a strong energy dependent bias, which we almost entirely eliminated (see figure \ref{fig:bias_vs_energy}), making our results more robust. The RNO-G collaboration has presented a shower energy reconstruction for a 'deep' station component as a function of neutrino energy \cite{energy_reco}. It is possible to compare the resolution for $\nu_e$ - CC events, as here the neutrino energy equals the shower energy, while considering the caveats that this is a different detector design embedded in Greenlandic ice and that analysis cuts had been applied in the RNO-G analysis. For  $E_{shower,\:pred} \: / \: E_{shower,\:true}$, RNO-G finds a resolution of \SI[parse-numbers = false]{1.1^{+1.1}_{-0.5}}{} at \SI{0.1}{\exa\eV} and \SI[parse-numbers = false]{1^{+0.5}_{-0.5}}{} at \SI{1}{\exa\eV}. With our neural network we find a resolution of \SI[parse-numbers = false]{1^{+0.3}_{-0.4}}{} both at \SI{0.1}{\exa\eV} and \SI{1}{\exa\eV} indicating a significant improvement especially at lower energies.

The better performance of the 'deep' station component can be traced back to the detector geometry. The 'shallow' station component uses 5 antennas spread over \SI{10}{\meter} vertically and \SI{6}{\meter} horizontally, while the 'deep' station component uses 16 antennas spread over \SI{100}{\meter} vertically and \SI{35}{\meter} horizontally. This gives the 'deep' station component a better map of the emitted Cherenkov cone, making it easier to reconstruct the vertex position and therefore the shower energy, even if several of the 16 antennas miss the signal. 

Previous analyses have shown that the signal strength in certain antennas has a large impact on the achievable resolution \cite{energy_reco,dir_reco}. To study this effect, we extract the signal-to-noise ratio (SNR) for every antenna and every event. For the signal strength, we determine the noiseless signal for every shower and ray tracing solution contributing to the event and save the maximum amplitude they reach in the recorded window. This approach made it possible to also quantify SNR strength below the noise level. To make it easier to see the discovered dependencies, we focus only on $\nu_x$ - NC events for the SNR study.

Regarding the reconstruction of shower energy in the 'shallow' component, we found that while the signal strength in the LPDAs is most important, the signal quality in the Vpol antenna also significantly affects the obtained resolution. The Vpol is sensitive to the vertical signal polarization, whereas the LPDA antennas are sensitive to the horizontal signal polarization. In the following, max(SNR$_{LPDA}$) denotes the strongest signal in any of the 4 LPDA antennas, while max(SNR$_{shallow \; Vpol}$) denotes the strongest signal in the Vpol antenna belonging to the 'shallow' station component. Figure \ref{fig:shallow_SNRvsE} shows how the obtained resolution relies on the SNR in the two antenna types. While a good signal strength in both is clearly best, at a low LPDA signal strength, the Vpol signal strength has a significant impact on improving the shower energy resolution. In the bottom plot, we further show how the resolution changes as a function of the shower energy for different Vpol SNR bins. The high-quality events above 3 SNR have a median uncertainty contour size of $\sim$0.1 in log(E), and these events make up about half of the triggers above a shower energy of $\SI[parse-numbers=false]{10^{18}}{\eV}$. This insight can help with the design of future in-ice radio neutrino detectors, to potentially deploy more than one Vpol per station if other constraints allow it.

\subsection{Direction Resolution}\label{sec: Results_direction}

\begin{figure*}[tbp]
  \centering
  \includegraphics[height=2.5in]{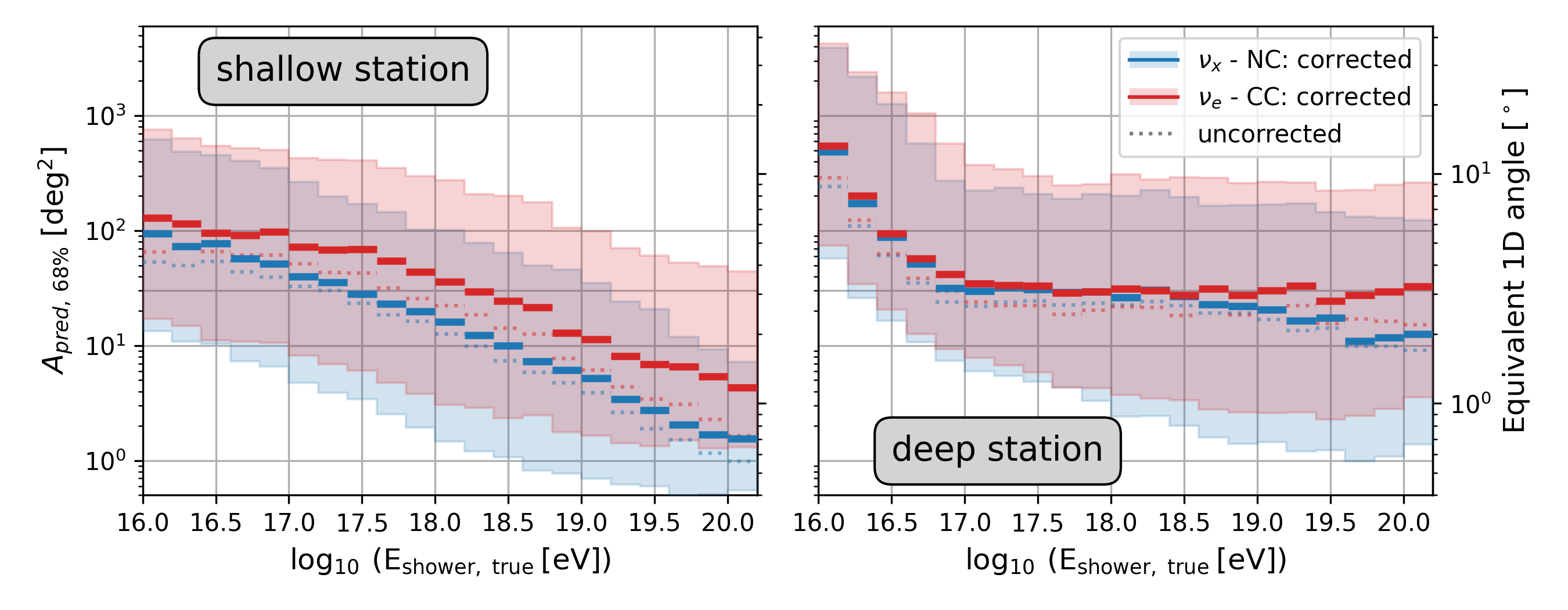}
  \caption{Results for the neutrino direction reconstruction on the test dataset. The median of the uncorrected resolution is indicated as a dotted line, while the median for the corrected resolution per energy bin is shown as a solid line. The shaded areas indicate the 16th percentile and the 84th percentile of the resolution per bin. The results for the '$\nu_x$ - NC' topology are blue while the results for the '$\nu_e$ - CC' topology are red. Left: Results for the 'shallow' station component. Right: Results for the 'deep' station component.}
  \label{fig:direction_resolution}
\end{figure*}

To determine the neutrino direction of an event, the neural network uses a 2-dimensional spherical spline normalizing flow. This allows us to model highly non-Gaussian uncertainty contours (these can commonly occur for in-ice radio neutrino detection, as explained in the introduction) on a 2-dimensional sphere, which can be parametrized by the azimuth and zenith angle. To evaluate the resolution of the direction reconstruction, we measure the size of the 2D 68\% contour of every event in the test dataset. To quantify the uncertainty using a single angle, as done in most previous analyses, we calculate the standard deviation of a Mieses-Fisher distribution of the same area. Also, here the uncertainty contour sizes were calculated by using the  68\% HDI, but now in two dimensions instead of one.

Figure \ref{fig:direction_resolution} shows the neutrino direction resolution on the test dataset as a function of the shower energy. For the 'shallow' station component, the reconstruction estimates a median 68\% uncertainty between 100 and 2 square degrees for $\nu_x$ - NC events and between 150 and 5 square degrees for $\nu_e$ - CC events. We observe an improvement in resolution with shower energy, as the direction reconstruction does not struggle with the same vertex distance dependence as the energy reconstruction. Also, for the neutrino direction reconstruction, it was easier for the model to reconstruct events of the $\nu_x$ - NC than the $\nu_e$ - CC event topology. The discrepancy in resolution between the two increases with shower energy as the effects of the LPM effect become more pronounced. When reconstructing $\nu_e$ - CC events, the model struggled more with under-coverage compared to $\nu_x$ - NC events, especially at high energies (see figure \ref{fig:comb_cov}, 10\% for $\nu_x$ - NC and 17\% for $\nu_e$ - CC events).

For the 'deep' station component, the reconstruction estimates a median uncertainty contour size between 500 and 10 square degrees for $\nu_x$ - NC events and between 600 and 40 square degrees for $\nu_e$ - CC events. For this station component, we observe a sharp improvement in resolution at low energies, and then the median resolution plateaus and only slightly improves further with shower energy. However, it is visible that the high-quality events in the 16th percentile continue to improve with increasing shower energy. This effect becomes even clearer when including the antenna SNR, where an increase in SNR of the triggering antenna always corresponds to a better resolution (seen in figure \ref{fig:deep_SNRvsE}). Also here, when reconstructing $\nu_e$ - CC events, the model struggled more with under-coverage compared to $\nu_x$ - NC events, especially at high energies (see figure \ref{fig:comb_cov}, 9\% for $\nu_x$ - NC and 14\% for $\nu_e$ - CC events).

For the 'shallow' station component, a neutrino direction resolution has previously been presented \cite{GLASER2023102781} for different neutrino energies. For $\nu_e$ - CC events, the results are directly comparable as shower energy equals neutrino energy. The previous study found a space-angle-difference for the 68th percentile of \SI[parse-numbers = false]{11}{\degrees}, \SI[parse-numbers = false]{6}{\degrees}, and \SI[parse-numbers = false]{4}{\degrees} at energies of \SI{0.1}{\exa\eV}, \SI{1}{\exa\eV}, and \SI{10}{\exa\eV} respectively. With our neural network we find a space-angle-difference for the 68th percentile of \SI[parse-numbers = false]{7}{\degrees}, \SI[parse-numbers = false]{5}{\degrees}, and \SI[parse-numbers = false]{3}{\degrees} at the same energies indicating a significant improvement especially at lower energies. For the 'deep' station components, a previous study found that 68\% of reconstructed events were contained in a contour of $\sim$1000 square degrees at \SI{1}{\exa\eV} \cite{dir_reco_POS}. Our analysis produces event-by-event uncertainty contours, making it possible to quantify the resolution based on the summary statistics of all tested events. At \SI{1}{\exa\eV} we observe a median contour size of $\sim$30 square degrees, indicating a significant improvement. 

When comparing the station components, it is important to consider their different behavior with shower energy. At $\SI[parse-numbers=false]{10^{17}}{\eV}$, the median resolution of the 'deep' station component is better than for the 'shallow' station component, as it has already reached its plateau, while the median resolution of the 'shallow' station component is still falling. At $\SI[parse-numbers=false]{10^{18}}{\eV}$ the resolution of the 'shallow' station component is better than for the 'deep' station component for $\nu_x$ - NC events but not for $\nu_e$ - CC events. At $\SI[parse-numbers=false]{10^{19}}{\eV}$, the resolution of the 'shallow' station component is better than the 'deep' component for both event types. Another difference between the two is that the high-quality $\nu_x$ - NC events in the 16th percentile have a consistently better resolution for the 'shallow' station component compared to the 'deep' station component, while for $\nu_e$ - CC events, the two station components perform very similarly.

\begin{figure}[tbp]
  \centering
  \includegraphics[height=8in]{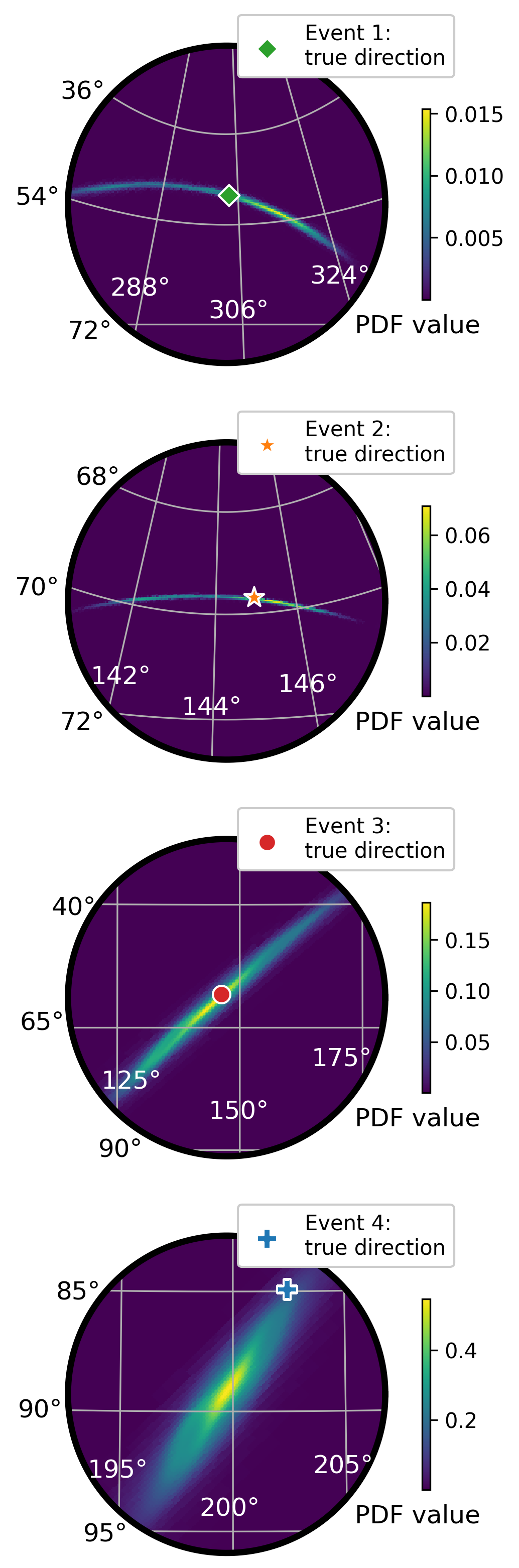}
  \caption{Posterior PDF predictions of the neutrino direction for 4 example events of the 'deep' detector component. The largest predicted contour is at the top, and they decrease in size until the bottom plot, which has the smallest contour size. All examples zoom in on different-sized windows, but the size of the uncertainty contour can be compared by the maximum PDF value in the color bars. }
  \label{fig:deep_skymaps}
\end{figure}

\begin{figure}[tbp]
  \centering
  \includegraphics[height=5in]{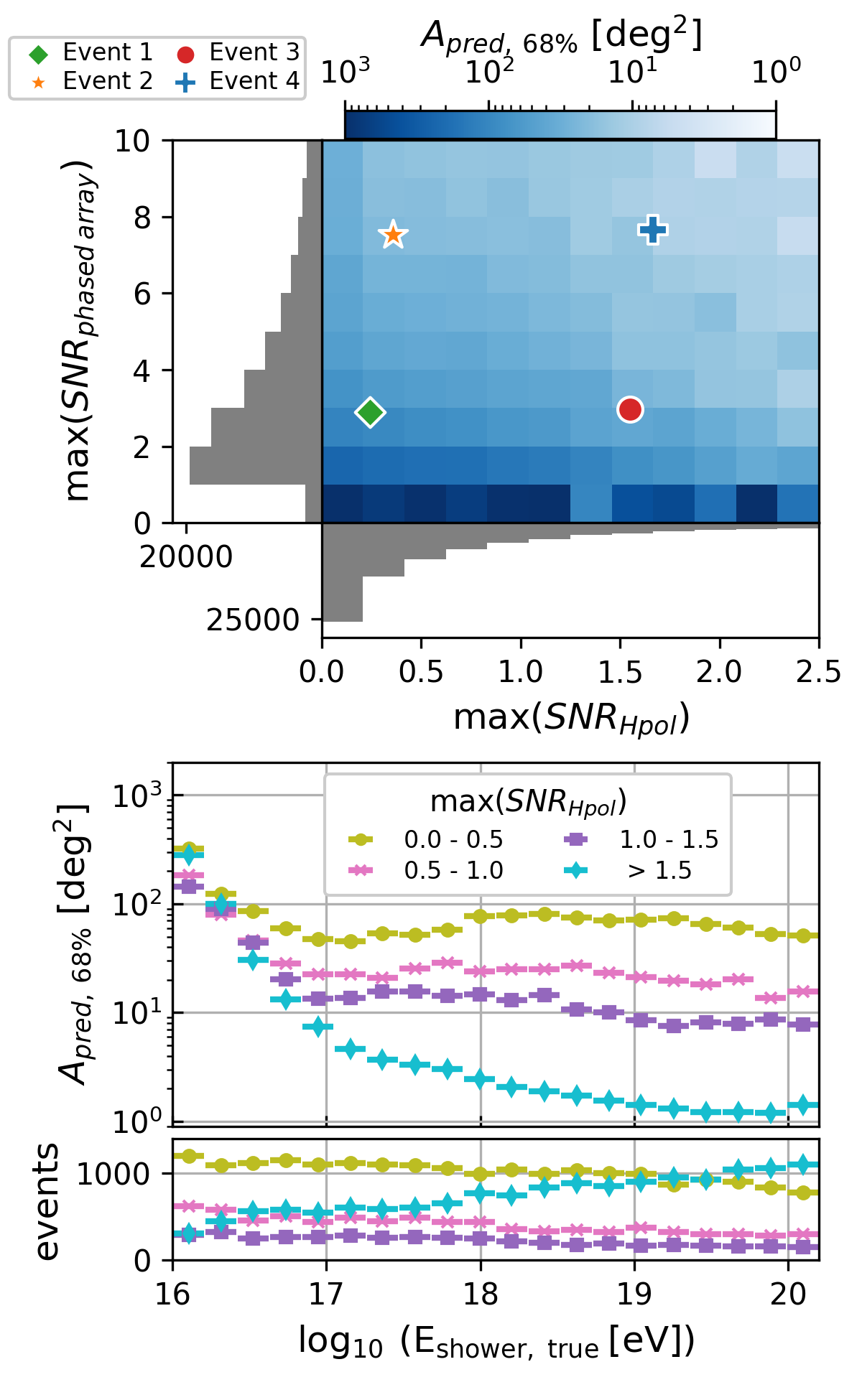}
  \caption{Impact of the 'deep' antenna SNR on the neutrino direction resolution for '$\nu_x$ - NC' events. Top: The maximum SNR of any of the ray tracing solutions in any of the 4 phased array antennas plotted against the maximum SNR of any of the ray tracing solutions in any of the 4 Hpol antennas. The color scale indicates the median neutrino direction resolution per bin, where a lighter shade of blue means a better resolution. The gray histograms indicate how many events are in each of the bins. However, this is not a physical SNR distribution as the data was generated in uniform energy bins. The 4 points on the plot correspond to the 4 example events plotted in figure \ref{fig:deep_skymaps}. Bottom: The corrected median neutrino direction resolution per shower energy bin for different Hpol SNR bins. The number of events is shown in the plot below.}
  \label{fig:deep_SNRvsE}
\end{figure}

Considering the shape of the uncertainty contours for the direction reconstruction gives further insight into which property of the signal helped the neural network in the reconstruction. Figure \ref{fig:overview} shows a typical uncertainty contour for the 'shallow' detector component, while figure \ref{fig:deep_skymaps} shows a variety of uncertainty contours for the 'deep' detector component. To quantify the asymmetry of the uncertainty contours, we calculate the median Kullback–Leibler(KL) divergence \cite{kl_div} of the events in the test dataset. For the 'shallow' detector component, we observe uncertainty contours with less elongated but thicker Cherenkov ring segments with a median KL-divergence of 0.4. For the 'deep' detector component, we often observe uncertainty contours with very elongated but thinner Cherenkov ring segments with a median KL-divergence of 2.3, indicating significantly more asymmetric contours. This can be traced back to the polarization and the viewing angle reconstruction. As a better polarization reconstruction limits the size of the ring segment from the Cherenkov cone, which is projected onto the sky, uncertainty contours with a better polarization reconstruction will be less elongated and more symmetric. This helps the 'shallow' detector component as the four LPDA antennas together with the Vpol antenna provide enough information to sufficiently reconstruct the polarization. As a better viewing angle reconstruction limits the thickness of the ring segment from the Cherenkov cone, which is projected onto the sky, uncertainty contours with a better viewing angle reconstruction will be thinner. This helps the 'deep' detector component, as the 12 Vpol antennas are spread out far enough to provide enough information to better reconstruct the viewing angle. These interpretable contour shapes also give us confidence that the neural network learned the underlying physics processes needed for the reconstruction instead of potential, nonphysical simulation artifacts. 

The Hpol antennas are sensitive to the horizontal signal polarization, making them a crucial component in the polarization reconstruction. To further study the extent of their impact on neutrino direction reconstruction, we compared the obtained resolution to the measured SNR in the Hpol antennas. In the following, max(SNR$_{phased \; array}$) denotes the strongest signal in any of the 4 Vpol antennas from the phased array, while max(SNR$_{Hpol}$) denotes the strongest signal in any of the Hpol antennas.

Figure \ref{fig:deep_SNRvsE} shows how the obtained resolution relies on the SNR in the two antenna types. While a good signal strength in both is clearly best, the Hpol signal strength has a significant impact on improving the neutrino direction resolution. In the bottom plot, we further show how the resolution changes as a function of the shower energy for different Hpol SNR bins. Here, it becomes clear how impactful a good signal strength in the Hpol antennas is for the neutrino direction resolution. The lowest SNR bin differs by a factor of 10 in contour size at low energies and almost by a factor of 100 at high energies compared to the highest SNR bin. To investigate this effect further, we retrained a model with only the 12 Vpol antennas (excluding all Hpols) but the same network architecture to see how the results would change. The median neutrino direction resolution was comparable to the 0.0 - 0.5 SNR bin, with a resolution above 40 square degrees for all shower energies. Interestingly, in the results of this retrained model, the best events in the 16th percentile also plateaued at $\sim$20 ($\sim$30) square degrees for the $\nu_x$ - NC ($\nu_e$ - CC) events. This is in stark contrast to the 16th percentile of the model, which included the Hpol antennas (figure \ref{fig:direction_resolution} and \ref{fig:deep_SNRvsE}), where the higher Hpol SNR events showed a significantly better neutrino direction resolution. This insight can help with the design of future in-ice radio neutrino detectors, to potentially deploy more Hpol antennas per station if other constraints allow it. 

The four events in figure \ref{fig:deep_skymaps} were chosen specifically such that their SNR is directly comparable with the other events (see figure \ref{fig:deep_SNRvsE} top for the SNR values). Event 1 has low SNR in both the phased array and the Hpol antennas, and its uncertainty contour spans about 75 degrees in azimuth and about 15 degrees in zenith. Event 2 (3) then shows how an improved phased array (Hpol) SNR impacts the shape. For event 2, the curved shape remains, with roughly the same extent in azimuth and zenith as event 1, but with a thinner contour band due to the higher SNR in the phased array antennas. Event 3, however, is only slightly curved and almost elliptical with a significantly smaller uncertainty band due to the higher SNR in the Hpol antennas. Event 4 is the smallest predicted contour with a very Gaussian shape as both the phased array antennas as well as the Hpol antennas have a high SNR.

When considering the Vpol antennas that were not included in the phased array of the 'deep' station component, we also observed a general improvement in resolution with increasing SNR. However, we noticed one interesting feature: the resolution worsened when the phased-array SNR was similar to the Vpol SNR at the higher antennas. This is mainly due to the viewing-angle reconstruction, which suffers when the amplitude distribution of the Cherenkov cone cannot be well mapped when two antennas show the same signal strength.

\subsection{Flavor Resolution}\label{sec: Results_flavor}

\begin{figure}[tbp]
  \centering
  \includegraphics[height=3in]{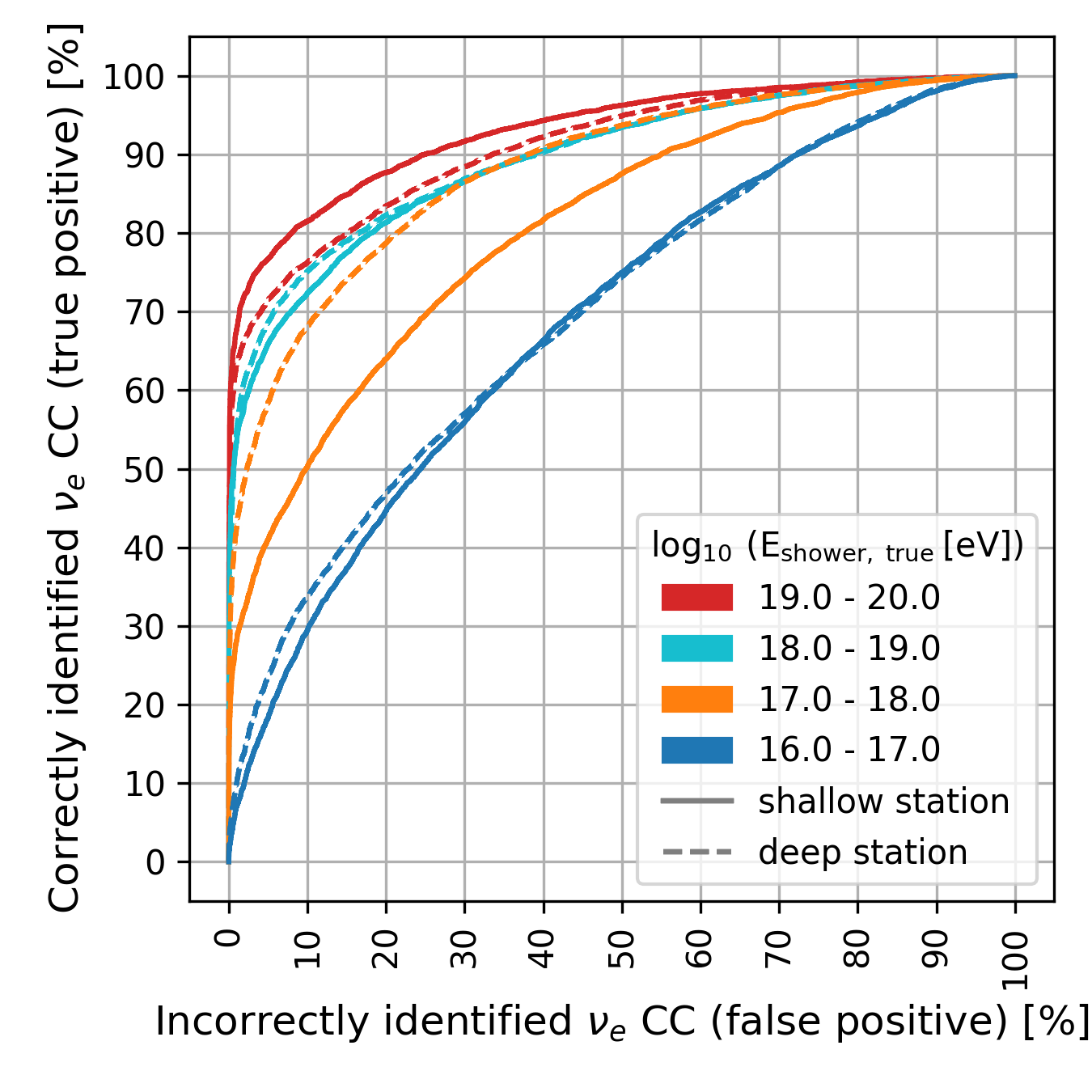}
  \caption{ROC-curve showing the results of the event topology classifier applied on the test dataset. The different colors show the energy bin of the events, and the line style indicates the station component with 'shallow' (solid) and 'deep' (dashed).}
  \label{fig:flavor_roc}
\end{figure}

As the sensitivity of a single station event to the flavor of the neutrino comes from deciding on the event topology of the shower, the neural network included a binary classification between $\nu_x$ - NC and $\nu_e$ - CC events. For every test event, a confidence between 0 and 1 is predicted (see example in figure \ref{fig:overview}, left/center). Depending on the analysis, the confidence threshold to decide between $\nu_x$ - NC and $\nu_e$ - CC events can vary. To retain this flexibility, we present the results in an ROC curve in figure \ref{fig:flavor_roc}, where different choices of thresholds result in different pairs of true positive and false positive rates. 

Even for very low-energy events where the LPM effect is significantly weaker, the model makes predictions better than a random guess. However, the higher the energy gets, the better the model predictions become as the LPM effect produces stronger variations in the $\nu_e$ - CC  showers. The two station components show different patterns in the ROC curves at different energies. The 'shallow' station component shows a constant improvement in classification as the energy increases. The 'deep' station component reaches a very good classification accuracy already at moderate energies, but does not improve as much beyond that. The performance for events with a shower energy in the ranges of $\SI[parse-numbers=false]{10^{16}}{\eV}$ - $\SI[parse-numbers=false]{10^{17}}{\eV}$ and $\SI[parse-numbers=false]{10^{18}}{\eV}$ - $\SI[parse-numbers=false]{10^{19}}{\eV}$ both the 'shallow' and 'deep' station components perform very similar. For events in the range of $\SI[parse-numbers=false]{10^{17}}{\eV}$ - $\SI[parse-numbers=false]{10^{18}}{\eV}$ the 'deep' station component performs better than the 'shallow' station component with up to 15\% lower false positive rates at the same true positive rate. For events in the range of $\SI[parse-numbers=false]{10^{19}}{\eV}$ - $\SI[parse-numbers=false]{10^{20}}{\eV}$ the 'shallow' station component performs better than the 'deep' station component with up to 10\% lower false positive rates at the same true positive rate.  

The results for the 'shallow station component are similar to previous results for this kind of station \cite{PhysRevD.110.023044}. For the 'deep' station component, it is the first time such a study has been performed.

\subsection{Systematic Uncertainties}\label{sec: Systematic}

The methodology of neural posterior estimation presented in this paper enables the inclusion of systematic uncertainties into the predicted posteriors \cite{thorsten_jammy}. This can be achieved by continuously sampling from all systematics when generating the training data set \cite{snowstorm}. Then the predicted posterior distributions will increase and will include both statistical and systematic uncertainties on an individual event-by-event level. Another advantage is that it is straightforward for the network to learn arbitrarily complex systematics, including correlations. However, quantifying systematic uncertainties is inherently difficult, especially for detectors that use natural media, in our case polar ice sheets, as active detector material. Furthermore, at the current development stage of in-ice radio detectors, systematic uncertainties have not yet been quantified. To still give an estimate of how systematics will impact the reconstruction, we study variations in the ice model and the antenna positions, which were identified as the likely most relevant systematic uncertainties in previous work \cite{dir_reco}. While the absolute values of the systematic errors are unknown, this study provides information on the qualitative dependence and informs calibration campaigns on the required level of precision. 

To estimate the impact of these systematics, we follow the standard approach and apply the trained model to new datasets with a constant variation in the simulation settings per dataset, in our case, different ice models and modified antenna positions or orientations. As these datasets differ from the data the neural network were trained on, this method provides the most pessimistic bound on the effects of systematic uncertainties, as neural networks often perform worse on unseen data. 

\begin{table*}[t]
\centering
\renewcommand{\arraystretch}{1.3} 

\begin{tabular}{l L{1.4cm} L{1.4cm} L{1.4cm} L{1.4cm} L{1.4cm} L{1.4cm} L{1.4cm} L{1.4cm}}

\toprule
\multirow{3}{*}{} 
& \multicolumn{4}{l}{'shallow' detector component} &\multicolumn{4}{l}{'deep' detector component}\\
\cmidrule(lr){2-5}\cmidrule(lr){6-9}
& \multicolumn{2}{l}{energy} & \multicolumn{2}{l}{direction}
& \multicolumn{2}{l}{energy} & \multicolumn{2}{l}{direction} \\
\cmidrule(lr){2-3}\cmidrule(lr){4-5}\cmidrule(lr){6-7}\cmidrule(lr){8-9}
& $\Delta$log(E) & coverage & $\Psi$[$^{\circ}$] & coverage
& $\Delta$log(E) & coverage & $\Psi$[$^{\circ}$] & coverage  \\
\midrule
baseline & \SI[parse-numbers = false]{-0.02^{+0.26}_{-0.30}}{} & -4\% & \SI[parse-numbers = false]{1.6^{+3.8}_{-1.1}}{} & -10\% & \SI[parse-numbers = false]{-0.01^{+0.11}_{-0.08}}{} & -5\%  & \SI[parse-numbers = false]{6.0^{+24.1}_{-5.0}}{} & -9\% \\
1\% ice  & \SI[parse-numbers = false]{0.20^{+0.23}_{-0.28}}{} & -34\%  & \SI[parse-numbers = false]{1.4^{+2.9}_{-1.0}}{} & -23\% & \SI[parse-numbers = false]{0.21^{+0.09}_{-0.11}}{} & -72\%  & \SI[parse-numbers = false]{3.9^{+17.5}_{-3.0}}{} & -21\% \\
5\% ice  & \SI[parse-numbers = false]{0.34^{+0.55}_{-0.36}}{} & -41\%  & \SI[parse-numbers = false]{6.1^{+38.1}_{-5.4}}{} & -47\% & \SI[parse-numbers = false]{0.21^{+0.09}_{-0.11}}{} & -72\%  & \SI[parse-numbers = false]{3.8^{+17.8}_{-3.0}}{} & -22\% \\
position & \SI[parse-numbers = false]{-0.07^{+0.41}_{-0.58}}{} & -30\%  & \SI[parse-numbers = false]{2.6^{+3.9}_{-1.5}}{} & -39\% & \SI[parse-numbers = false]{-0.01^{+0.09}_{-0.08}}{} & -7\%  & \SI[parse-numbers = false]{5.3^{+23.2}_{-3.0}}{} & -9\% \\
orientation & \SI[parse-numbers = false]{-0.02^{+0.26}_{-0.28}}{} & -5\%  & \SI[parse-numbers = false]{2.1^{+3.3}_{-1.4}}{} & -19\%  \\
\bottomrule
\end{tabular}

\caption{Results from the study of systematic uncertainties for the 'shallow' and 'deep' detector components for $\nu_x$ - NC events at \SI[parse-numbers=false]{10^{18}}{\eV}. The values for baseline dataset were taken from figure \ref{fig:bias_vs_energy} for the shower energy bias and the space-angle difference, and from figure \ref{fig:comb_cov} for the maximum under-coverage. Shown is the median value where the subscript is the 16th percentile and the superscript is the 84th percentile.}
\label{tab:systematics}
\end{table*}

Each systematics dataset includes 5.000 re-simulated $\nu_x$ - NC events per shower energy bin at \SI[parse-numbers=false]{10^{17}}{\eV}, \SI[parse-numbers=false]{10^{18}}{\eV}, and \SI[parse-numbers=false]{10^{19}}{\eV}. The altered datasets were generated by re-simulating triggered events from the original simulated dataset that were not included in the training or testing of the models. This introduces a small bias to higher SNR events in the altered datasets, but this bias is estimated to be negligible. It also means that the 5000 events are not exactly the same events for each systematic as some events trigger for one systematic but not for the other. However, there remains a large overlap of events in the datasets. In this way, we can calculate the impact of single systematic changes but disregard potential correlations between several simultaneous changes. To further verify that a resimulation of already triggered events does not introduce a bias, we also resimulated 5000 events without any changes to the simulation settings. 

We changed the ice model parameters (of a two-parameter exponential model) in one dataset by 1\% and in a second by 5\%.  For the 'shallow' detector component, we also considered a change in the antenna's horizontal and vertical positions, sampled from a Gaussian with a standard deviation of \SI{10}{\centi\meter}, as well as a change in antenna orientation sampled from a Gaussian with a standard deviation of 5 degrees. For the 'deep' station component, we did not treat the antenna positions as independent variables but considered vertical and horizontal shifts by string. We sampled the string positions from a Gaussian with a standard deviation of \SI{10}{\centi\meter} for the vertical position and \SI{5}{\centi\meter} for the horizontal position. 

As expected, we find that the network under-predicts the uncertainties by evaluating the coverage. Hence, we use the space-angle difference and energy difference between the predicted and MC true values to estimate the reconstruction resolution. 
The results of the systematic uncertainty estimation can be seen in table \ref{tab:systematics}. There was a small but not significant reduction in shower energy bias and space-angle difference with a rising shower energy, which is why we focus on the uncertainty values at \SI[parse-numbers=false]{10^{18}}{\eV}. Some of the values presented showed a slightly smaller shower energy bias and space-angle difference compared to the baseline dataset. We interpret this small effect as a consequence of the parameter in question not being significant to alter the result, limited statistical precision, and the slightly higher SNR in the systematics datasets, improving the resolution. For the 'shallow' detector component, we see that the 1\% change in ice model impacted the coverage of the predicted contours and significantly shifted the shower energy predictions. With the 5\% change in ice model parameters, we see an increase in this effect, and also a significant shift in the space-angle difference. The shift in antenna position also showed significant under-coverage but only a minor effect on the shower energy bias and the space angle difference. Only minor effects can be observed in the results from the rotated antenna orientation dataset. For the 'deep' detector component, the ice model had a large impact on the shower energy prediction with a large under-coverage and bias compared to the baseline dataset. This is due to the small size of the predicted uncertainty contours, leading to a large under-coverage when the true value is shifted slightly. In the direction reconstruction, we also see a significant rise in under-coverage but not an associated worsening of the space-angle difference, which is made possible by the asymmetric uncertainty contours. The shift in antenna position showed minor to no difference compared to the baseline model.

It is clear that systematic uncertainties, especially those from the ice parameters, can have a major impact on the reconstruction quality and must therefore be measured and treated carefully. Both station components show significant worsening in the energy reconstruction when the ice parameters change. Changes in the antenna position mostly impacted the 'shallow' station components, while changes in the angular orientation of the antennas only had a minor effect.

\section{Goodness-of-Fit}\label{sec: Goodness}

A reconstruction based on a neural network, as presented in this work, will always produce an output (shower energy, neutrino direction, event topology) no matter what input it receives, as long as the input has the correct dimensions. We therefore studied how the model can filter out data that it is unfamiliar with. To filter out signals from background sources such as anthropogenic noise, cosmic ray events, or wind-induced signals, we developed a goodness-of-fit check that calculates how similar a newly measured event is to the Monte Carlo data with which the neural network was trained \cite{thorsten_jammy}. In addition, such a method can help in verifying the simulated signal against a measured neutrino signal in the future. This method is explained in the following paragraph.

\begin{figure}[tbp]
  \centering
  \includegraphics[height=4in]{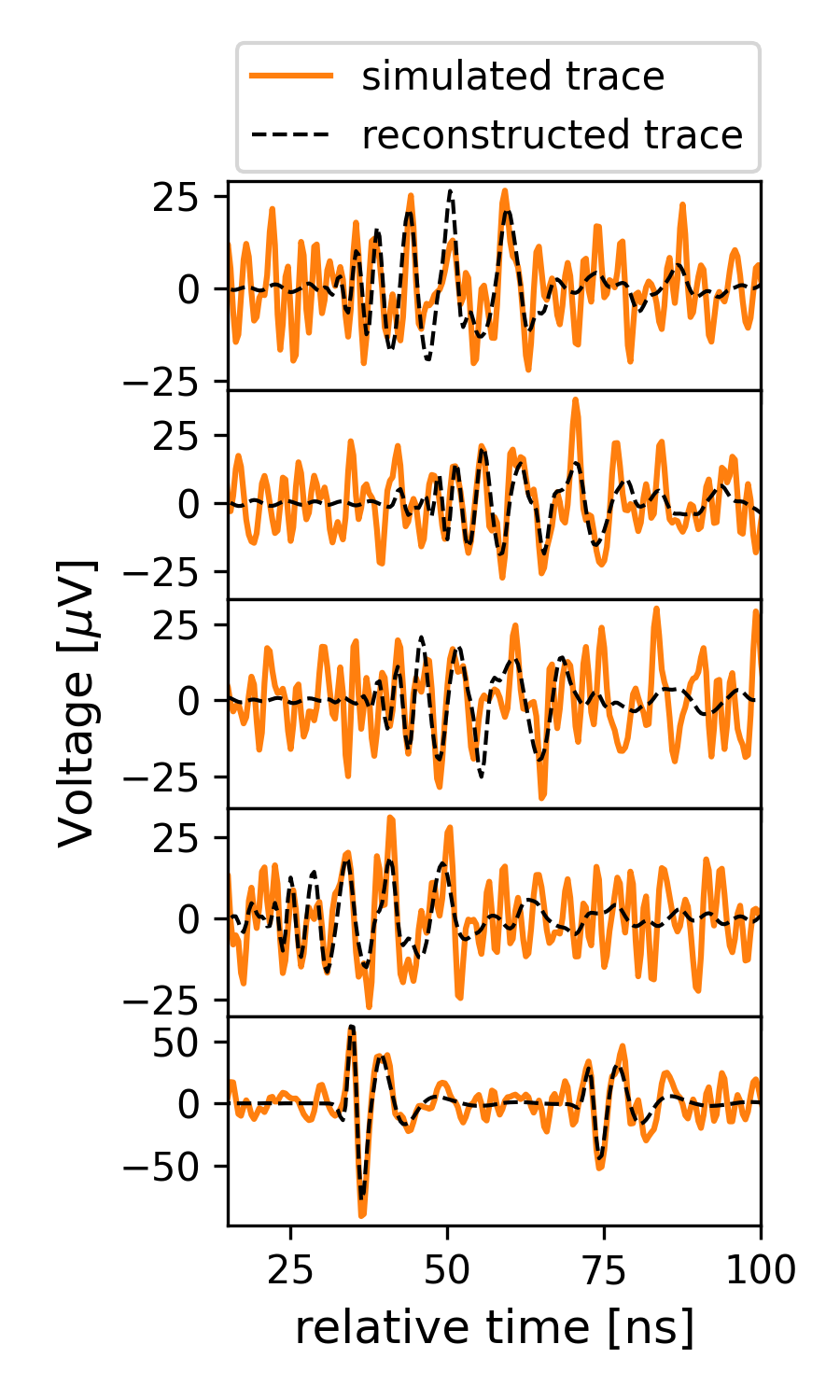}
  \caption{Example neutrino signal from the test dataset of the 'shallow' detector component, with the best-fit signal prediction from the model prediction in black. The four top plots correspond to the 4 LPDA antennas, while the bottom plot corresponds to the Vpol antenna. The signal region has been cut out from the full data event. This is the same event as shown in figure \ref{fig:overview}}
  \label{fig:shallow_trace}
\end{figure}

\begin{figure}[tbp]
  \centering
  \includegraphics[height=2.2in]{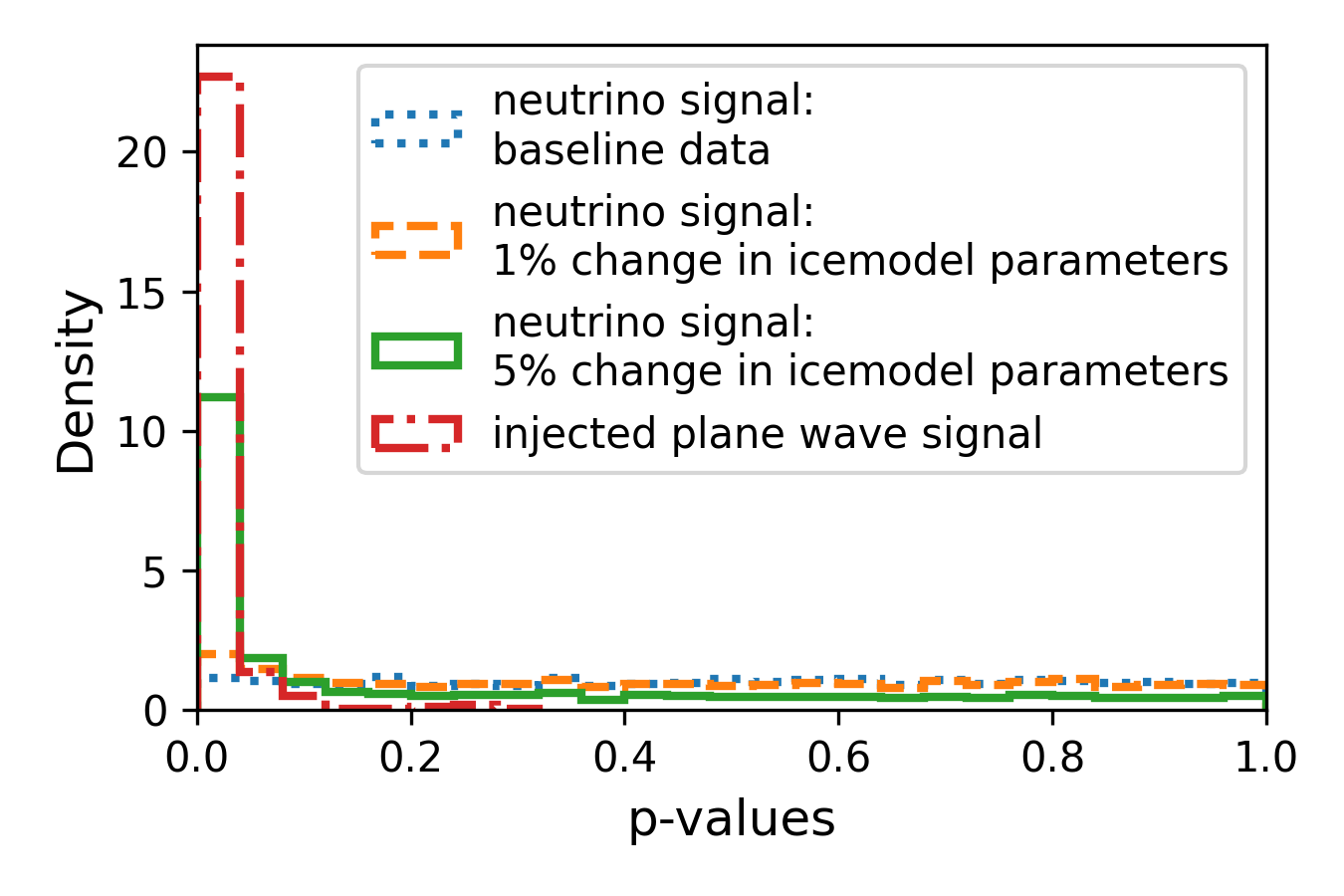}
  \caption{P value distribution derived from the goodness-of-fit score for different datasets. The blue dotted line shows the uniform distribution for events from the 'baseline' dataset. Orange and green show the distributions for events with changed ice model parameters and red shows the dataset with injected plane wave signals.}
  \label{fig:p_values}
\end{figure}

\begin{figure*}[tbp]
  \centering
  \includegraphics[height=5in]{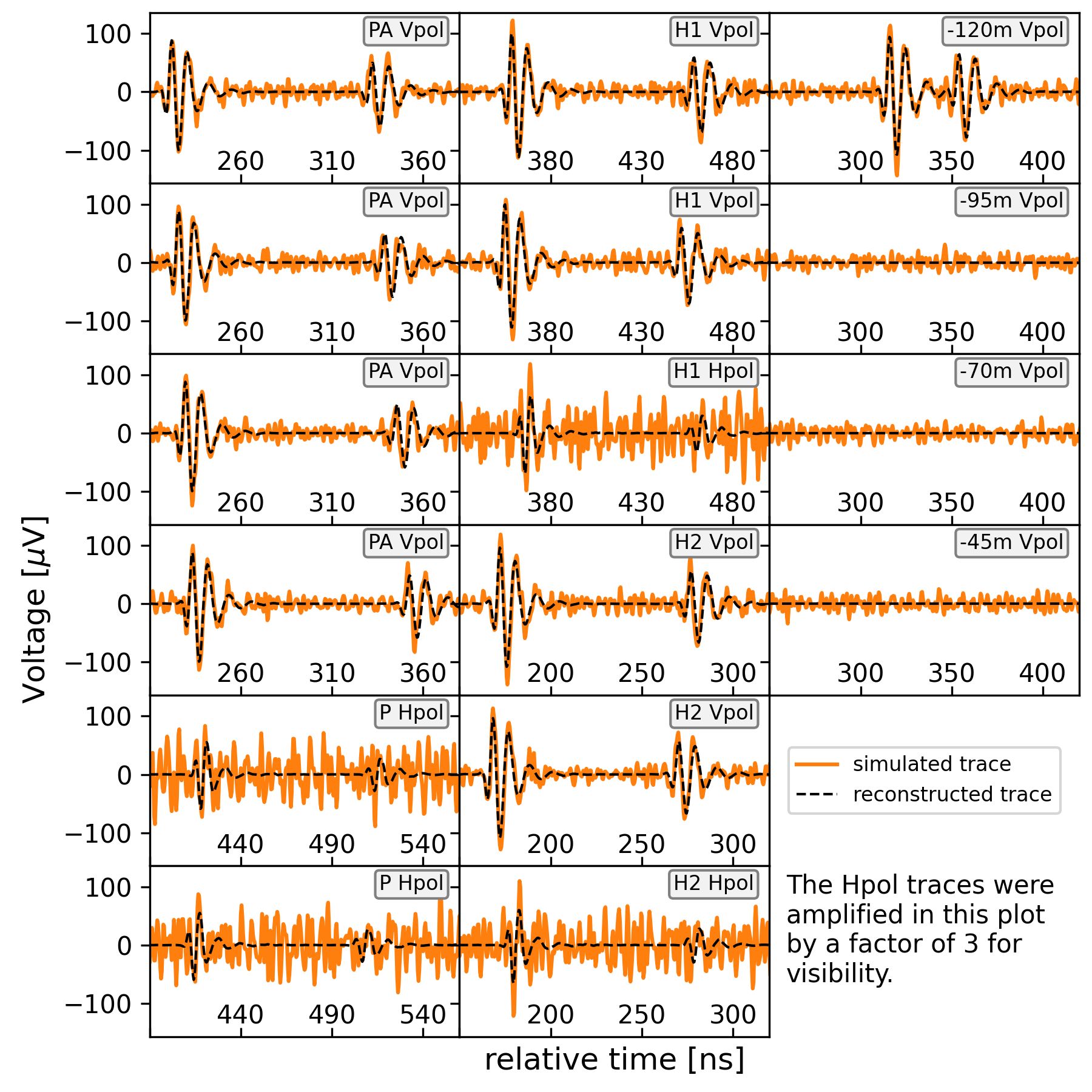}
  \caption{Example neutrino signal from the test dataset of the 'deep' detector component, with the best-fit signal prediction from the model prediction in black. The left column corresponds to the 4 phased array antennas (PA Vpol) and the two Hpol antennas right above the phased array (P Hpol). The center column corresponds to the antennas on the two helper strings (H1 and H2), and the right column corresponds to the 4 Vpol antennas, which are at different heights on the power string. The signal region has been cut out from the full data event.}
  \label{fig:deep_trace}
\end{figure*}

A simulated neutrino signal relies on 7 parameters. The shower energy (one parameter), the neutrino direction (two parameters), the position of the interaction vertex (three parameters), and the absolute event time (one parameter). The first six parameters are all predicted by the presented neural network, where we use the mode of the predicted PDFs as the best estimate values for all parameters. These values are used, together with the simulation specification of the detector and ice description, to predict noiseless voltage signals as they would reach the detector. As this does not include any trigger simulation, the absolute time is fitted by comparing the noiseless signal to the noisy event of interest using a normalized cross-correlation. This is done simultaneously across all antennas to fit a single offset time. For simplicity, this test was only performed for $\nu_x$ - NC events, as the shower realizations for these events only have minor variations, which are fitted as well. After these steps, we obtain the best estimate noiseless trace for every event in the 'baseline' dataset, of which an example can be seen in figure \ref{fig:shallow_trace} for the 'shallow' detector component and in figure  \ref{fig:deep_trace} for the 'deep' detector component. Recently, a likelihood description for radio detectors was developed \cite{Ravn:2024arx} which allows for a statistically interpretable comparison between the noiseless signal and the noisy test trace by calculating the $-2 \Delta$ log-likelihood between them. For the Monte Carlo true parameters of shower energy, neutrino direction, and vertex position, the calculated score follows a chi-squared distribution where the degrees of freedom correspond to the number of time samples in the event. For numerical stability, frequencies with less than 10\% of the maximum amplitude were ignored. In this way, we can treat the calculated $-2 \Delta$ log-likelihood values as a test statistic. For the 'baseline' events in our test set, which were generated with the same MC settings as the training dataset, we calculate this score and the distribution they produce. For every 'new' event that we want to probe, we can calculate a p-value of how likely it is that this event comes from the same distribution. For new events from the 'baseline' test dataset, the calculated p-values follow a uniform distribution between 0 and 1 (figure \ref{fig:p_values}, blue). For events that differ from the data the model was trained on, the p-value distribution will be skewed towards 0, and the larger the discrepancy, the stronger the skew will be. 

This method was then applied to different datasets. First, we simulated 500 plane wave events (mimicking other noise sources like anthropogenic noise) with a similar SNR spectrum as the neutrino events with varying polarizations and arrival directions. All the filters and trigger conditions were identical to the neutrino simulation. We then applied our deep learning model to these events and calculated the goodness-of-fit score as well as the p-values for each of them. The results can be seen in figure \ref{fig:p_values}, where the red line corresponds to the plane wave signals. We can reject many of them with high confidence, as almost all of them have a p-value close to 0. Second, we also calculated the goodness-of-fit score for the datasets from the study of the systematic uncertainties, where especially the changes in ice models showed a large shower energy bias and under-coverage. As the calculation of the goodness-of-fit score assumes the nominal ice model parameters, we suspected it would be possible to filter them out in a similar way as the plane wave signals. The 5\% change in the ice parameters results in a strongly skewed distribution. Many events pile up at p values of 0, but also a significant number of events are still visible at high p values (figure \ref{fig:p_values}, green).  For a 1\% change in ice parameters, the distribution is only slightly skewed, and it will be difficult to identify such a small data-MC disagreement (figure \ref{fig:p_values}, orange).
Overall, this approach offers a novel method for verifying the compatibility of the MC simulations on which the neural network was trained with measured data. The higher the signal-to-noise ratio of an event, the better this method will work.

The highest rate of background events for in-ice radio neutrino detectors will come from thermal noise fluctuations. We saw that the models were successful in identifying these events even without the goodness-of-fit score, as they produced very large uncertainty contours, especially for the direction reconstruction. We simulated 1000 thermal noise fluctuations that pass the trigger condition, with the same filter and trigger configurations as the neutrino simulations, and saw a median uncertainty size of \SI[parse-numbers = false]{7200^{+2800}_{-4100}}{} square degrees for the 'shallow' station component and \SI[parse-numbers = false]{4800^{+1700}_{-1600}}{} square degrees for the 'deep' station component. This indicates that the models are successful in filtering out low-SNR thermal noise fluctuations through large uncertainty contours and impulsive non-neutrino signals, such as plane waves, through the goodness-of-fit score, where discrimination improves with increasing event SNR.

\section{Summary and Outlook}\label{sec: Conclusion}

We developed a neural network capable of performing a full reconstruction for in-ice radio neutrino detectors. For the first time, we apply neural posterior estimation to in-ice radio detection, combining a deep neural network with conditional normalizing flows to predict the posterior distribution for each parameter of interest on an event-by-event basis. This is of particular interest for in-ice radio neutrino detectors due to the expected asymmetric uncertainties in the neutrino direction.
Furthermore, neural posterior estimation provides a unique way to include systematic uncertainties directly in the predicted uncertainty contours. This will become important in the future, once a better understanding of the underlying systematic uncertainties is reached. Finally, we introduced a goodness-of-fit score that enables testing the consistency between the neural network predictions and the observed data. This metric allows us to identify events that are inconsistent with a neutrino origin (e.g., anthropogenic background) as well as significant discrepancies between the underlying Monte Carlo simulations and reality (e.g., deviations in the ice model).

We analyzed the proposed IceCube-Gen2 radio detector, including its 'shallow' and 'deep' detector components, in terms of shower energy and neutrino direction resolution, as well as flavor separation capability. We found significant improvements compared to previous analyses without any quality cuts applied to the data. In particular, a strong energy-dependent bias, previously observed in the shower energy predictions of the 'shallow' detector components, was eliminated. For the 'deep' detector components, the shower energy resolution was improved by roughly a factor of two without applying analysis cuts, and the neutrino direction indicates an improvement in resolution by roughly a factor of 30 in the uncertainty area. 
We note that the network’s ability to predict individual uncertainties enables straightforward application of quality cuts, e.g., to select a subset of events with low directional uncertainty.

Furthermore, to guide future detector optimizations, we studied the impact of the different antennas on the reconstruction performance. We found that the signal strength in the Vpol antenna of the 'shallow' component has a strong positive effect on the energy resolution. Similarly, we found that the angular resolution of the 'deep' component crucially depends on the signal strength in the Hpol antennas. Overall, the predicted uncertainties behave consistently with theoretical expectations, indicating that the network has learned the correct physical dependencies.

\section*{Acknowledgments}
This work is supported by the European Union (ERC, NuRadioOpt, 101116890) and by the Swedish Research Council (VR) via the project 2021-05449.

\bibliographystyle{JHEP}
\bibliography{ref}

\appendix

\section{Coverage}\label{sec: Coverage}

\begin{figure}[tbp]
  \centering
  \includegraphics[height=3.5in]{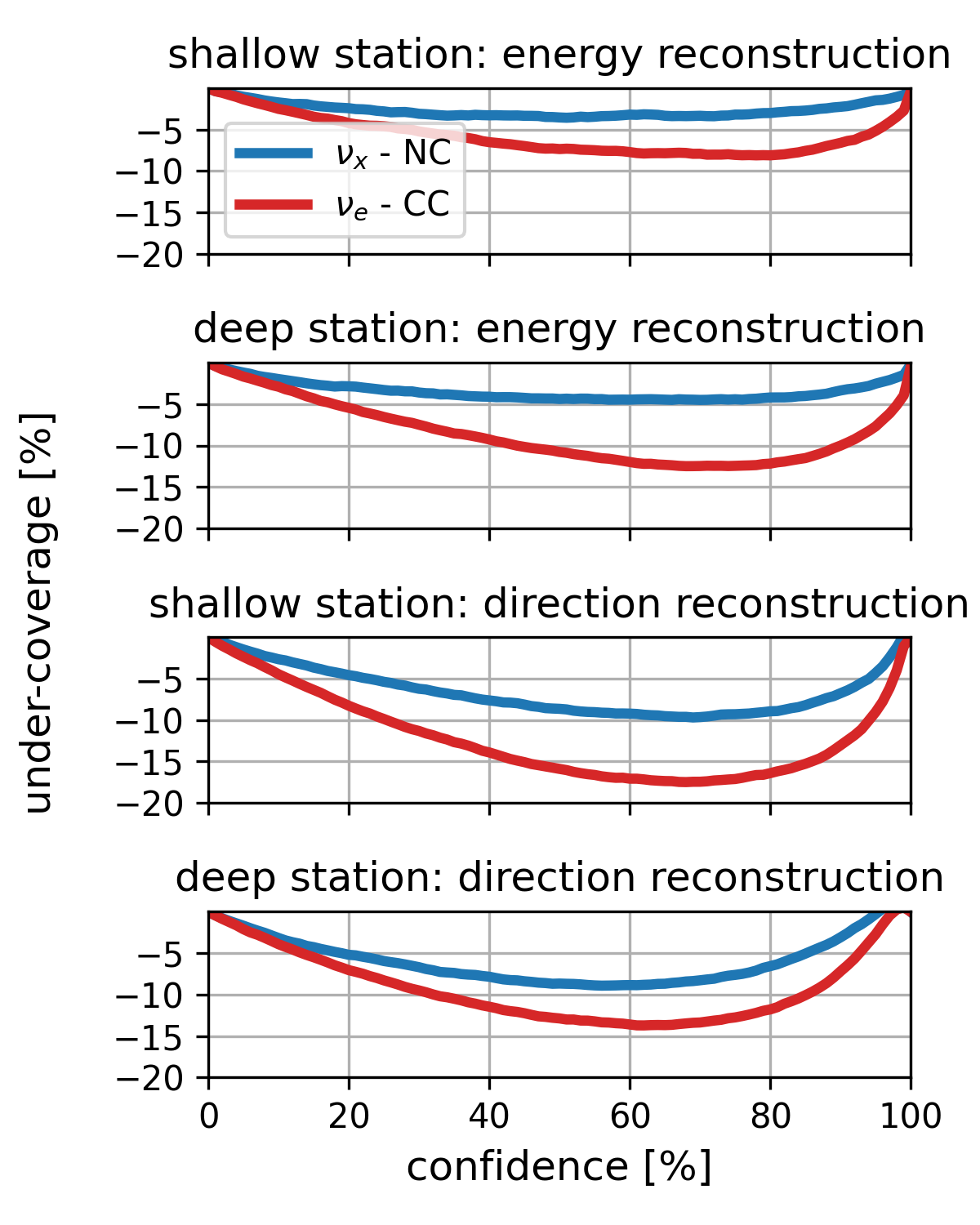}
  \caption{Coverage for the energy and direction reconstruction for the 'shallow' and 'deep' components. The data was split by event topology with $\nu_x$ - NC in blue and $\nu_e$ - CC in red.}
  \label{fig:comb_cov}
\end{figure}

Figure \ref{fig:comb_cov} shows the coverage for each reconstruction. For all of them, we observe a non-negligible under-coverage, which originates from balancing the resolution and accuracy when finding the correct moment to stop the training. As the under-coverage was significantly lower when tested on the training dataset, it seems to be an artifact from over-training the model. However, when comparing the final corrected resolution to different steps during training, the presented model still had the best results. This is also not an effect introduced because of the convoluted loss-function, as the same under-coverage appears when only training on reconstructing energy/direction or when only training on $\nu_x$ - NC / $\nu_e$ - CC events. Although work will continue to reduce this under-coverage, the correction in the figures \ref{fig:energy_resolution} and \ref{fig:direction_resolution} shows that the effect is small enough that we still obtain results with an excellent resolution after taking coverage into account.

\section{Correlation of Energy and Vertex Position}\label{sec: Correlation}

\begin{figure*}[tbp]
  \centering
  \includegraphics[height=4.1in]{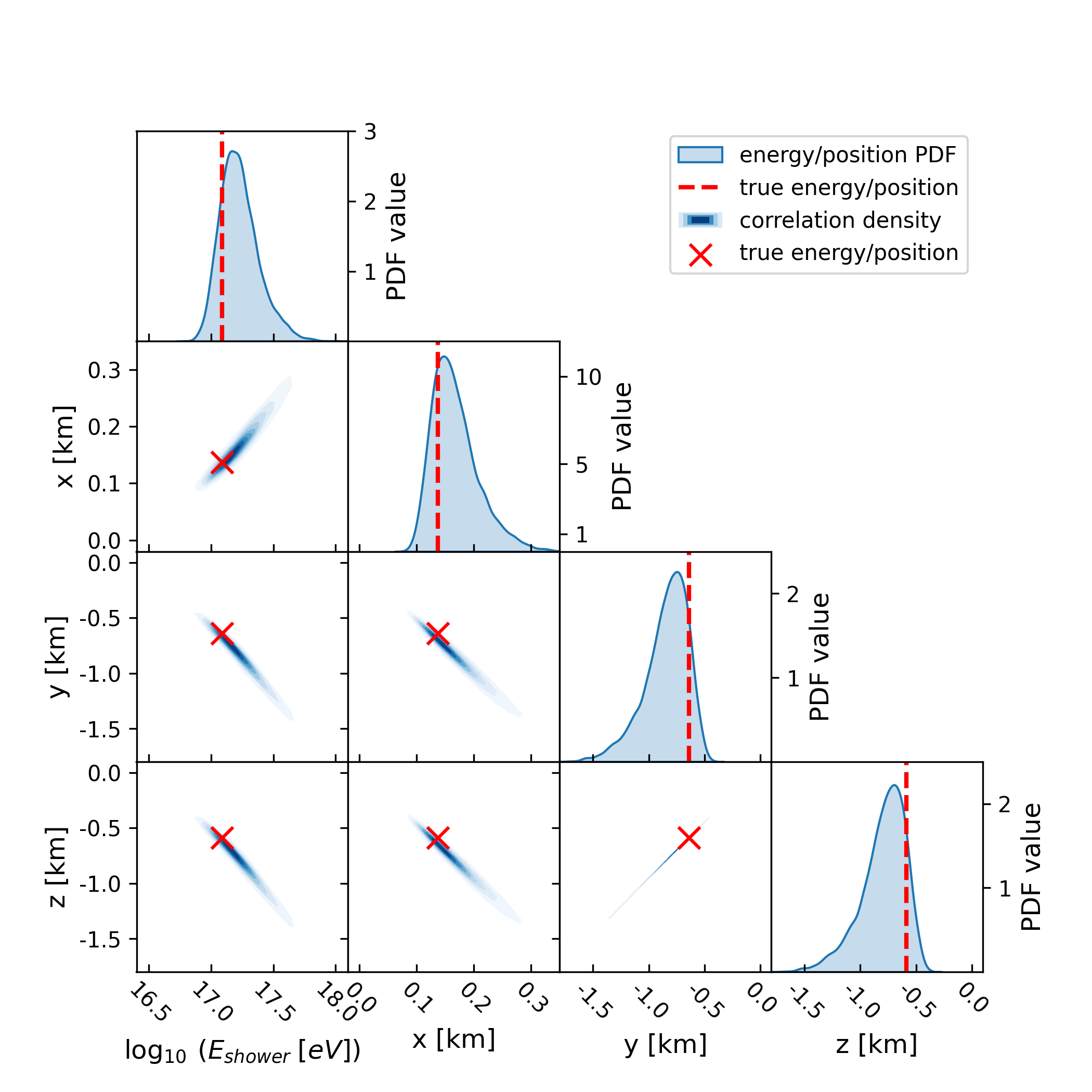}
  \caption{Corner plot to show the correlations between the shower energy and vertex position predictions for the example event shown in figure \ref{fig:overview} and \ref{fig:shallow_trace}. The rightmost plot in each row shows the marginalized PDF for each dimension, and the 2D-histograms show the correlations between the different dimensions.}
  \label{fig:correlations}
\end{figure*}

As our neural network predicts the energy and the vertex position together in a 4-dimensional PDF, it's possible to analyze the correlations between them on an event-by-event basis. An example of this can be seen in figure \ref{fig:correlations}. It becomes clear that the correlations across the four dimensions are very strong, meaning that the network learned that if the energy is lower than its best estimate, the vertex position must have been closer to the detector to produce the same signal. The same is true for the correlations between the vertex position parameters.

\section{Bias and Space-Angle Difference}\label{sec: bias}

Even though we are able to quantify the size of the event-by-event uncertainty contours, it is interesting to consider the metrics used in previous analyses to quantify the quality of reconstruction. 

\begin{figure*}[tbp]
  \centering
  \includegraphics[height=3.7in]{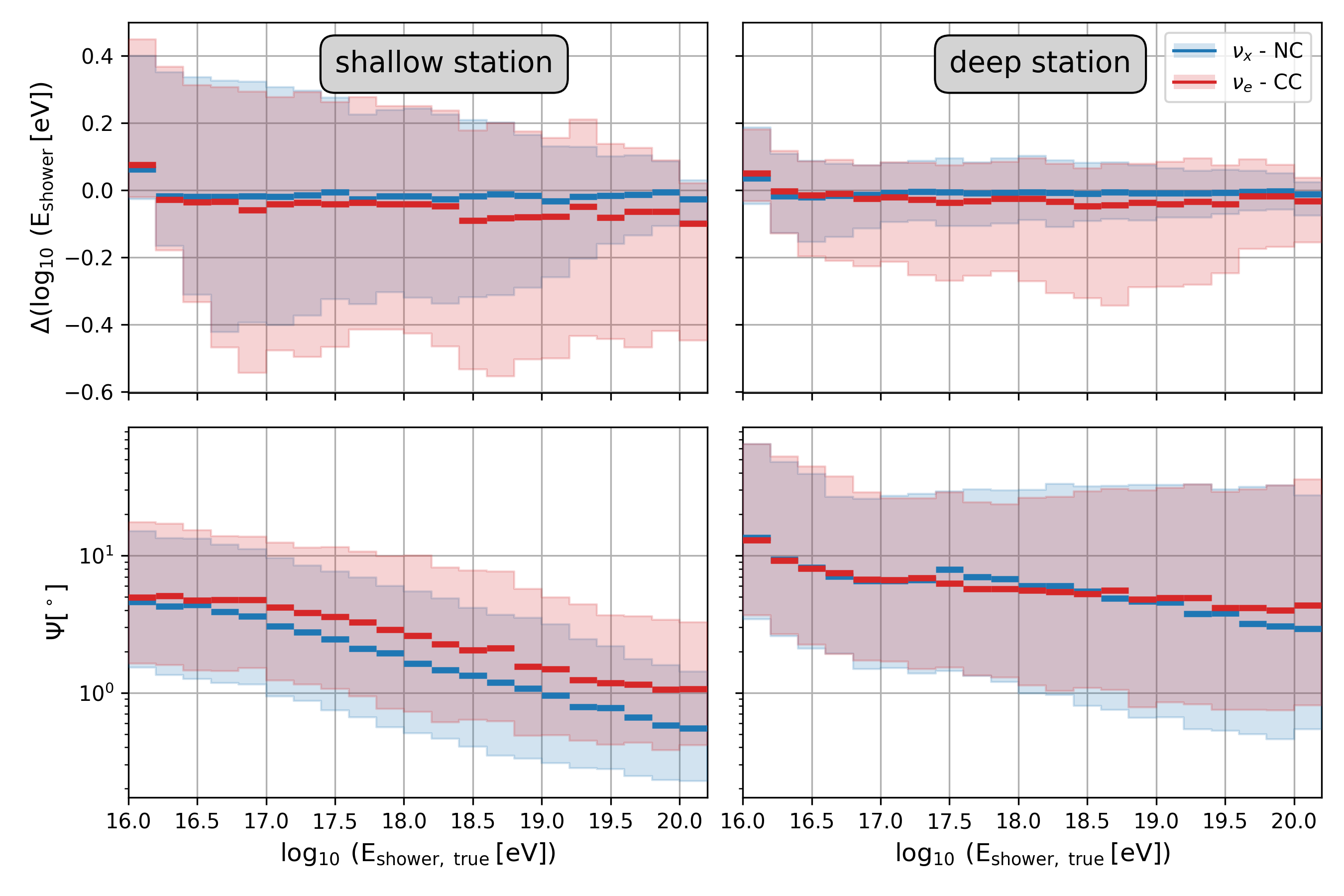}
  \caption{Resolution in terms of energy bias and space-angle-difference. The lines show the values of the median, while the shaded region shows the 16th and 84th percentiles of the distribution. Top: Bias plots for the shower energy prediction, where the y-axis shows the difference between the predicted (taken from the mode of the predicted PDFs) and the MC true shower energy. Bottom: Space-angle difference between the predicted (taken from the mode of the predicted PDFs) and the MC true direction vector.}
  \label{fig:bias_vs_energy}
\end{figure*}

The energy bias is the difference between the 'true' shower energy from the MC simulations and the best-estimate shower energy from the reconstruction. In our case, the best estimate of the shower energy is extracted by taking the mode from the marginalized shower energy PDF. The results can be seen in figure \ref{fig:bias_vs_energy} (top). Both station components show a median energy bias centered around 0 with minor deviations at low energies. Similar to the size of the uncertainty contours, the 'deep' station component shows a lower shower energy bias in for the region between the 16th and 84th percentile compared to the 'shallow' station component. Also, the $\nu_e$ - CC events show a significantly stronger bias compared to the $\nu_x$ - NC events, especially at higher energies.

The space-angle difference is the angle between the MC true direction vector and the reconstructed neutrino direction vector. Again, we extract the best reconstructed vector from the mode of the spherical PDF trained on the neutrino direction. Here, we see a very similar behavior with shower energy for both station components compared to the size of the uncertainty contours. Notably, the space-angle difference for the 'deep' station component is higher than estimated from the size of the predicted uncertainty contours, hinting at highly asymmetric contour shapes as seen in figure \ref{fig:deep_skymaps}.

\end{document}